\documentclass[%
twocolumn,  
secnumarabic,
amssymb, 
nobibnotes,
aps, 
prl,
superscriptaddress,
nofootinbib,
amsmath,amssymb
]{revtex4-1}

\usepackage{graphicx}
\usepackage{dcolumn}
\usepackage[mathlines]{lineno}
\usepackage{amsmath}
\usepackage{array}
\usepackage{hyperref}
\hypersetup{
    colorlinks=true,
    linkcolor= [rgb]{0.2,0.8,0.5},
    citecolor= [rgb]{0.1,0.7,0.1},
    filecolor=magenta,      
    urlcolor=[rgb]{0.1,0.5,0.5},
}
\usepackage{cleveref}
\usepackage{float}

\newcommand*\aap{Astron.~Astrophys.}

\newcommand*\apjl{Astrophys.~J.~Lett}

\newcommand*\pasp{PASP}

\newcommand{\LCDM}{$\Lambda$CDM }
\newcommand{\Hzero}{H_0}
\newcommand{\Mpc}{{\rm  Mpc}}
\newcommand{\diff}{\rm d}
\newcommand{\Omzero}{\Omega_{\rm m 0}}
\newcommand{\wde}{w_{\rm de}}
\newcommand{\rhom}{\rho_{\rm m}}
\newcommand{\rhode}{\rho_{\rm de}}

\newcommand{\Panp}{Pan$^+$}

\graphicspath{{./}{Figures/}}

\begin{document}

\title{Inferring dark energy properties from the scale factor parametrisation }


\author{Upala Mukhopadhyay}
\email{rs.umukhopadhyay@jmi.ac.in}
\affiliation{Centre For Theoretical Physics, Jamia Millia Islamia, New Delhi, 110025, India}
\author{Sandeep Haridasu}%
\email{sharidas@sissa.it}
\affiliation{SISSA-International School for Advanced Studies, Via Bonomea 265, 34136 Trieste, Italy}
\affiliation{INFN, Sezione di Trieste, Via Valerio 2, I-34127 Trieste, Italy}
\affiliation{IFPU, Institute for Fundamental Physics of the Universe, via Beirut 2, 34151 Trieste, Italy}
\author{Anjan A Sen}%
\email{aasen@jmi.ac.in}
\affiliation{Centre For Theoretical Physics, Jamia Millia Islamia, New Delhi, 110025, India}%
\author{Suhail Dhawan}%
\email{suhail.dhawan@ast.cam.ac.uk}
\affiliation{Institute of Astronomy and Kavli Institute for Cosmology, University of Cambridge, Madingley Road, Cambridge CB30HA,UK}%

\begin{abstract}

We propose and implement a novel test to assess deviations from well-established concordance $\Lambda$CDM cosmology while inferring dark energy properties. In contrast to the commonly implemented parametric forms of the dark energy equation-of-state (EoS), we test the validity of the cosmological constant on the more fundamental scale factor [$a(t)$] which determines the expansion rate of the Universe. We constrain our extended `general model' using the late-time observables. The posterior of the dark energy EoS is mainly constrained to be quintessence-like naturally excluding physically unviable regions such as phantom crossings or exponential growth. 

\end{abstract}

\maketitle
\section{Introduction}\label{sec:Introduction}
Over the last couple of decades, increased accuracy and precision of the cosmological observables have provided strong support for a $\Lambda$CDM model \citep{Peebles2024} since the first hints in \cite{Riess98}. This cosmology has been extremely successful in explaining a wide range of phenomena, including but not limited to the fluctuations in the temperature and polarisation of the cosmic microwave background (CMB) \cite{Planck18_parameters, Planck2020, Planck15_DE, ACT:2020gnv}, observations of the large-scale structure of the universe \cite{Aubourg_2015, Alam16, Addison:2017fdm, Haridasu17_bao}, as well as the distance-redshift relation of Type Ia supernovae \cite{Riess:2019qba, Scolnic:2017caz, Brout:2022vxf}. However, more recent observations of the baryon acoustic oscillations from DESI survey \cite{DESI:2024mwx}, combined with Type Ia supernovae \cite{Rubin:2023ovl, DES:2024tys} and the CMB hint at deviations from the canonical $\Lambda$CDM model at $\sim 3-4 \sigma$ level. Moreover, with increased precision, the Cepheid-calibrated local distance ladder measurement of the Hubble Constant  \citep[
$H_0$;][]{Riess2022} is in $> 5\sigma$ tension with the inference from the CMB \citep{Planck2020}. These observations, along with theoretical problems, e.g. the elusive nature of the dark matter particle and the cosmological fine-tuning problem \citep[e.g., see][]{Weinberg2000}, motivate the need to explore alternate cosmologies.

Exploring alternatives to the concordance \LCDM model can be divided into main scenarios: (i) Deviation from the $\Lambda$, i.e. $\wde = -1$ scenario, where $\wde$ is the equation of state (EoS) of dark energy (DE). This can either be explored via a simple extension to a time-invariant  $\wde \neq -1$, or with parametric forms that include a time-dependence of $\wde(z)$. The most common is $\wde(a) = w_{0} + (1-a) w_{a}$ - popularly known as CPL parameterisation \cite{CHEVALLIER_2001, Linder_2003} - where $w_0$ and $w_a$ are the present day value of the equation of state and its derivative wrt scale factor, respectively. This model while being phenomenological, encompasses a large family of dynamical DE models \cite{Linder:2005in, Linder02, Capozziello06}.  In this scenario, it is assumed that Einstein's General Relativity (GR) is the correct theory of gravity at cosmological scales and that the universe at late times is dominated by a pressure-less matter component with EoS $w_{\rm m}=0$, which consists of dark matter(DM) and baryons; (ii) a scenario where there is no DE in the Universe and the Universe consists of pressure-less matter but the late time acceleration of the Universe is driven by a modified version of gravity at large cosmological scales. Examples of such modified gravity are $f(R)$ gravity theories \cite{Sotiriou_2010, Nojiri:2017ncd}, scalar-tensor gravity \cite{YasunoriFujii_2003}, Cardassian models \cite{Freese_2002}, DGP  model \cite{Dvali:2000hr}, Modified Chaplygin models \cite{Barreiro_2004} etc; (iii) a less explored option is when the DE is given by a Cosmological Constant with $w_{de} = -1$ but the matter sector is not entirely pressure-less, $w_{m} \neq 0$ \cite[see also \cite{Colgain:2022tql}]{Vattis:2019efj, Clark20, Haridasu:2020xaa, FrancoAbellan:2020xnr, Naidoo:2022rda, Poulot:2024sex, Lapi:2023plb}. 

In this study, we are instead, interested in the most fundamental quantity of the cosmological background evolution, namely the scale factor, $a(t)$ within the FLRW Universe. All the observable related to background evolution are constructed from this fundamental quantity. In fact, all the cosmological observations related to the geometry of the background Universe can only constrain the time derivative of the scale factor $H = {\dot{a}}/{a}$ and is not sensitive to how $H(t)$ is decomposed into different components of the Universe or the type of modified gravity as long as they produce the same expansion history \cite{Linder_2003}. Moreover, at sub-horizon scales where perturbation in the DE sector can be neglected, the growth in matter fluctuations solely depends on the $H(t)$ of the Universe and the total matter content ($\Omzero$) \cite{Linder:2005in, Zhumabek:2023ejd}. This is also true for modified gravity theories whereas the late Universe only contains the matter sector and the large-scale cosmic acceleration is driven by the modified dependence of $H(t)$ on the matter-energy density $\rhom(t)$. 

Given this, we use low redshift cosmological observations, e.g the data from SNIa observations, from BAO observations as well as growth data from measurements of matter fluctuations at linear scale to constrain the evolution of $a(t)$ or $H(t)$. Note that our formalism to test deviations from \LCDM is valid only at the late-times and hence appropriately, the late-time observables. Once we constrain the $a(t)$ (and $H(t)$), assuming that low redshift Universe is dominated by a minimally coupled dark energy together with pressure-less dust ($\Omzero$), the constraints can be easily extended to study the nature of dark energy. We intend the current analysis as a pilot study to be extended to a full-fledged assessment of the more fundamental scale-factor evolution. 


\section{Modelling}\label{sec:model}

Within the standard \LCDM model, the exact analytical form of the scale factor at late times can be written as, 
\begin{equation}
\label{eqn:sf_LCDM}
a(t) = a_0^{1/3} [\sinh (t/\tau)]^{2/3}.
\end{equation}
where the contribution of radiation/relativistic species is assumed to be negligible at late times and $\tau^2 = \frac{4}{3 \Lambda c^2}$ \footnote{In terms of the fractional densities one can write $\tau = 2/(\sqrt{\Omega_{\Lambda 0} 3 \Hzero)}$. Also, the parameter $a_0 = \frac{8 \pi G \rhom}{\Lambda} \equiv \frac{\Omzero}{\Omega_{\Lambda 0}}$ can be straight away related to the fractional densities within the \LCDM model.} \cite{Reid:2002kp, Gron:2002wrd} is an arbitrary parameter having the dimension of time [{\it T }]. This results in the form of the standard Hubble parameter $H(z)$ as, 
\begin{equation}
\label{eqn:H_LCDM}
H^{2} (z) = H_{0}^{2} [\Omzero (1+z)^3 + \Omega_{\Lambda 0}].
\end{equation}
where the fractional density parameters $\Omega_{i} = \frac{\rho_i}{3\Hzero^2/ 8\pi G}$. Note that in deriving the \cref{eqn:sf_LCDM}, one has to assume the closure equation $\Omega_{\Lambda 0} = 1- \Omzero$, assuming flatness ($\Omega_{\rm k} = 0$)\footnote{See \cite{Anselmi:2022uvj, Jimenez:2022asc}, for more recent discussion on the assumption of flatness in comsological modelling.}, and the solution cannot be immediately extended to the radiation-dominated epoch. This form of the scale factor $a(t)$ and the subsequent form for the Hubble parameter $H(z)$ are sufficient to estimate any observable for the background Universe in DE-dominated epochs. Moreover, as discussed in the Introduction, the information for $a(t)$ or $H(z)$ is the only cosmological information needed apart from the information on $\Omzero$ and the normalization constant $\sigma_{8}$ to calculate the observable related to the growth of matter perturbations \cite{Linder:2005in}, at late times.
\subsection{Model extension}\label{sec:ABmodel}
Given that we have established the form of the $a(t)$ in \cref{eqn:sf_LCDM} corresponding to the standard \LCDM scenario, we now describe the extended formalism. We study the possible deviations to the same, by constraining the behaviour of scale factor in a minimally extended formalism. To this end, we introduce an extension to \cref{eqn:sf_LCDM}, to add flexibility to the $a(t)$ in the following way \cite{Sen_2002}:

\begin{equation}\label{eqn:sf_ABCDM}
a(t) = a_{1}^{2/B} [\sinh (t/\tau)]^{B},
\end{equation}
where $a_{1} $ and $B$ are dimensionless arbitrary parameters and $\tau$ is again an arbitrary parameter of dimension [{\it T }]. Here a value of $B \neq 2/3$  indicates deviation from the standard \LCDM scenario. As shown in \cite{Sen_2002}, a value of $B<2/3$ gives a slower descent to acceleration and vice-versa for $B>2/3$. With this, one can calculate $H(t) = \frac{\dot{a}}{a}$ and finally replacing  $a(t)$ by redshift $1/(1+z)$, one can arrive the expression for the Hubble parameter $H(z)$:

\begin{equation}\label{eqn:H_ABCDM}
H^{2} (z) {\equiv \Hzero^2 E^2(z)} = H_{0}^{2} [A (1+z)^{2/B} + (1-A)].
\end{equation}
Here $A = \frac{a_{1}^{2/B}}{1+a_{1}^{2/B}}$ and $H_{0}^2 = \frac{B^2}{\tau^2 (1-A)}$ is the Hubble parameter at present. In the rest of the paper, we mention this model as ``General Model". Note that the above \cref{eqn:sf_ABCDM,eqn:sf_ABCDM} describes the deviation from \LCDM model without any specific assumptions for DE behaviour. In comparison to the \LCDM model, it has one extra parameter. For observable related to background evolution only, this form of the deviation from \LCDM is completely agnostic to whether late time acceleration is driven by a DE or by a late time modification of gravity at large scales. However, all the deviations are now modelled through the index $B$, which drives the evolution. With $B=2/3$ and with the identification $A = \Omega_{m0}$, we retrieve the $H(z)$ for \LCDM model. One can easily derive the expression for the `dimensionless' age of the Universe $t_{0}$ for our model given by \cref{eqn:sf_ABCDM} as, 
\begin{equation}\label{eqn:age}
    t_{0} H_{0} = \frac{B}{\sqrt{1-A}} \rm{Sinh}^{-1} \left[ \sqrt{\frac{1-A}{A}}\right]
\end{equation}

Once again, with $B=2/3$ and $A=\Omzero$, this expression reduces to the corresponding expression for the \LCDM model. It is interesting to note that observables like the expansion rate of the Universe at present ($\Hzero$), the age of the Universe ($t_{0}$) etc, are analytically related to the parameters that appeared in the scale factor $a(t)$ without any assumption on the energy content of the Universe.

\noindent
There may be several interpretation of $H(z)$ behaviour given in \cref{eqn:H_ABCDM}:

i) Due to a DE which scales as:
\begin{align}\label{eqn:rhode}
\rhode(z) &\sim A(1+z)^{2/B} + (1-A)\nonumber\\
&- \Omzero (1+z)^3, 
\end{align}

\noindent
together with a pressure-less matter which scales normally as $\sim \Omzero(1+z)^{3}$. If $B=2/3$ and $A=\Omzero$, then $\rhode$ is a constant, and the model behaviour reduces to \LCDM model. If $B=2/3$ and $A \neq \Omzero$, then $\rhode\neq$ constant and model is yet not the \LCDM model but is an evolving DE model, wherein the evolving part of DE behaves tracks the matter density. For $B \neq 2/3$, it is clearly an evolving DE model.

ii) Due to modification of Einsteins' GR at late times at cosmological scales with matter as the only density contribution and $H^{2} \propto \rhom^{3/B} + c$, where $c$ is a constant. This is similar to the Cardassian model \cite{Freese_2002} for modified gravity. 

iii) Due to matter which may not be entirely pressure-less and scales as $\sim (1+z)^{2/B}$ instead of $\sim (1+z)^3$ as in pressure-less case. In this case, one should interpret the parameter $A$ straightway as $\Omzero$.  

iv) Due to a scenario where a pressure-less matter is interacting with an evolving  DE however still with $w_{\rm de} = -1$, such a scenario has been previously studied in the context of the Generalized Chaplygin Gas model \cite{Bento_2004}.

The first interpretation is the simplest one and in this case, one can also use the data from the perturbed Universe e.g, the growth data to constrain the $\Omzero$ (together with $A$ and $B$) and can subsequently reconstruct the DE equation of state $\omega_{\rm de}(z)$. Moreover, there may be an interesting scenario in this case: As one can see from \cref{eqn:age} if $B > 2/3$, the term proportional to $(1+z)^{2/B}$ will scale slower than $(1+z)^3$ at high redshift and hence $\rhode$ may be negative at higher redshift for certain values of $A$, although the total energy density of the Universe will always be positive. A similar scenario in the context of ``Omnipotent Dark Energy'' has been recently studied in \cite{Adil_2024}. Finally as shown in \cite{Sahni:2015hbf}, the $H(z)$ behaviour shown in \cref{eqn:sf_ABCDM} can also be obtained with a K-Essence field $\Phi$ rolling over a constant potential.

\section{Data}\label{sec:data}

We consider well-established and most recent cosmological datasets to constrain the parameters of the `general model'. We enlist the datasets here: \textit{i)} We have considered the Baryon Acoustic Oscillation (BAO) distance and the correlated expansion rate measurements from the Sloan Digital Sky Survey (SDSS)-IV collaboration. We have used 14 BAO measurements compiled in Table 3 of Ref.~\cite{eBOSS:2020yzd}. Hereafter we term this dataset as `SDSS'. More recently the Dark Energy Spectroscopic Instrument (DESI) observation has presented robust measurements of BAO within similar redshift ranges as the SDSS compilation. We have taken 12 BAO measurements assembled in Table 1 of Ref.~\cite{DESI:2024mwx}. We term them `DESI' hereafter. Note that we utilise these two compilations interchangeably to provide information from the BAO observables and contrast the constraints from either. \textit{ii)}: We have used the type Ia supernovae (SNe Ia) distance moduli measurements from the Pantheon+ sample \cite{Brout:2022vxf}. It consists of 1590 distinct SNe Ia in the redshift range z $\in$ [0.01, 2.26]. Hereon, we represent this dataset as \Panp. \textit{iii)}: In a late Universe, where we have a minimally coupled DE field together with a pressure-less matter component, the evolution of matter density contrast $\delta$ neglecting fluctuations in the DE-filed can provide constraints on matter density ($\Omzero$). For growth data, we have considered Table 3 of Ref. \cite{Nesseris:2017vor} where the updated and extended growth-rate data are provided in terms of the $f\sigma_{8}$ parameter at different redshifts where $f = \frac{\diff \log(\delta)}{\diff \log(a)}$ and $\sigma_{8}$ is the variance of matter density fluctuations in a sphere of comoving radius $8 \, h^{-1}$ Mpc. We should mention that when we use the growth data, we assume that the $H(z)$ expressed in \cref{eqn:H_ABCDM} is represented as a minimally coupled dark energy model with $\rhode(z)$ given by \cref{eqn:rhode} together with an ordinary pressureless matter ($p_{m} = 0$). 

In our model, we have the following parameters, $\{A, B\}$ as the model parameters together with $h\times r_{\rm d} \equiv \alpha$ (appearing in BAO observables) and the absolute magnitude ($M_{\rm b}$) of SNIa, when included in the analysis. Here the present-day Hubble parameter $H_{0} = 100\, h$ Km/s/Mpc and $r_{\rm d}$ sound horizon at drag epoch. Also, when we use the growth data assuming that $H(z)$ in \cref{eqn:H_ABCDM} along with $\rhode(z)$ which evolves as \cref{eqn:rhode}, we have two additional cosmological parameters $\Omzero$ and $\sigma_{8}$. In this case, one can also reconstruct the equation of DE $\omega(z)$ as,

\begin{equation}
\omega(z) = \frac{\omega_{\rm T}(z) E^{2} (z)}{E^{2}(z) - \Omzero (1+z)^3},\label{EOS}
\end{equation}
where the total EoS of the Universe $\omega_{\rm T}(z) = \frac{p_{\rm T}(z)}{\rho_{\rm T}(z)}$  is given by
\begin{equation}
    \omega_{\rm T}(z) = \frac{2A}{3B}\frac{(1+z)^{2/B}}{A(1+z)^{2/B} + (1-A)} -1,\label{EOST}
\end{equation}

\noindent
and $E^2(z)$ is given by \cref{eqn:H_ABCDM}. Finally, we estimate $w_0$($\equiv w(z=0)$) and $w_a$($\equiv \frac{{\diff} w(z)}{{\diff} z}\mid_{z=0}$) for our model to compare it with the CPL type DE model around present day ($z=0$). {Note that the CPL-based estimation of the dynamical DE behaviour includes parameter space that need not always be physically viable \cite{Peirone:2017lgi, Raveri:2017qvt} (see also \cite{ Vagnozzi:2018jhn}). These include theoretical considerations such as the Cauchy problem, exponential growth and Ghost condition. }

Finally, we write down a simple Gaussian joint likelihood for the three datasets summarized above and perform a Bayesian analysis, assuming generous uniform priors on all parameters. {We utilise the publicly available \texttt{EMCEE} \footnote{Available at: \href{http://dfm.io/emcee/current/}{http://dfm.io/emcee/current/}} package \citep{Foreman-Mackey13}} to perform the sampling and \texttt{GetDist} \footnote{Availalbe at: \href{https://getdist.readthedocs.io/}{https://getdist.readthedocs.io/}} \cite{Lewis:2019xzd} to analyse the chains. We also compute the Bayesian Evidence \cite{Trotta:2017wnx, Trotta:2008qt} to assess the model-selection utilising \texttt{MultiNest}\footnote{Availalbe at: \href{https://github.com/JohannesBuchner/MultiNest}{https://github.com/JohannesBuchner/MultiNest}} \cite{Feroz:2008xx}.

\section{Results}\label{sec:resutls}

In \cref{triangle} we show the constrained parameter space, alongside the point $B=2/3$ and $A=0.315$ (Planck-2018 \cite{Planck18_parameters}) which when put in the expression for $H(z)$ in \cref{eqn:H_ABCDM} reduces to the best fit $\Lambda$CDM model. We find good agreement with the CMB constraints only when considering SNIa and BAO (DESI) datasets. As one combines the growth data, albeit weak, one gets an additional constraint on matter density $\Omzero$ and unless the constraints on $A$ and $\Omzero$ are consistent with each other, even with $B = 2/3$, the model does not reduce to $\Lambda$CDM model as in the expression of $\rhode$ in \cref{eqn:rhode}, the first and last term do not cancel each other and hence $\rhode$ is evolving and can not reduce to a cosmological constant.

From \cref{triangle}, our primary inferences for the constraints on $\{A, B\}$ parameter space can be summarized as follows:
Using only Pantheon+ data, the values $B=2/3$ and $A = 0.315$ are allowed within $\sim 1\sigma$ showing no strong departure from Planck-\LCDM behaviour. Moreover, a large section of allowed parameter space falls in the $B > 2/3$ region showing the possibility of negative DE behaviour at larger redshift. Using only SDSS BAO data, the point $B=2/3$ and $A = 0.315$ are only at the boundary of the $\sim 2\sigma$ allowed region, whereas, for the more recent DESI compilation, \LCDM point is within $1\sigma$ region. This shows that with our parametrization in \cref{eqn:sf_ABCDM} for the scale factor $a(t)$, Planck-$\Lambda$CDM is more consistent with DESI data than SDSS although for both, Planck-$\Lambda$CDM is allowed within $\sim 2\sigma$. Moreover, the allowed parameter space for DESI is much tighter than SDSS showing the better constraining power of the former than the latter. {Clearly, this mild difference is driven by the data points $z \lesssim 0.7$ within both the BAO compilations, as noted in \cite{DESI:2024mwx}. A comparison of the constraints in the CPL method indicates the same, as in, considering only BAO data only\footnote{Inclusion of the \Panp data makes no difference to this inference on the $\{w_0, w_a\}$ parameter space. Note also that the $>3\sigma$ inference for the deviation form \LCDM in \cite{DESI:2024mwx}, is strongly driven by the inclusion of more recent SNIa compilation in \cite{DES:2024tys}, which is not present when the earlier \Panp is used.}, DESI compilation is in better agreement with \LCDM than the earlier SDSS data. } When we add Pantheon+ to SDSS or DESI BAO datasets, we find mild shifts of the allowed parameter space towards the $B>2/3$ region. For SDSS+\Panp, the $B=2/3$ and $A = 0.315$ is allowed only at $\sim 2\sigma$ whereas, for DESI+\Panp, it is allowed at $1\sigma$ showing lesser deviation and no hints for new physics beyond Planck-$\Lambda$CDM model.

\begin{figure*}
\includegraphics[scale = 0.38]{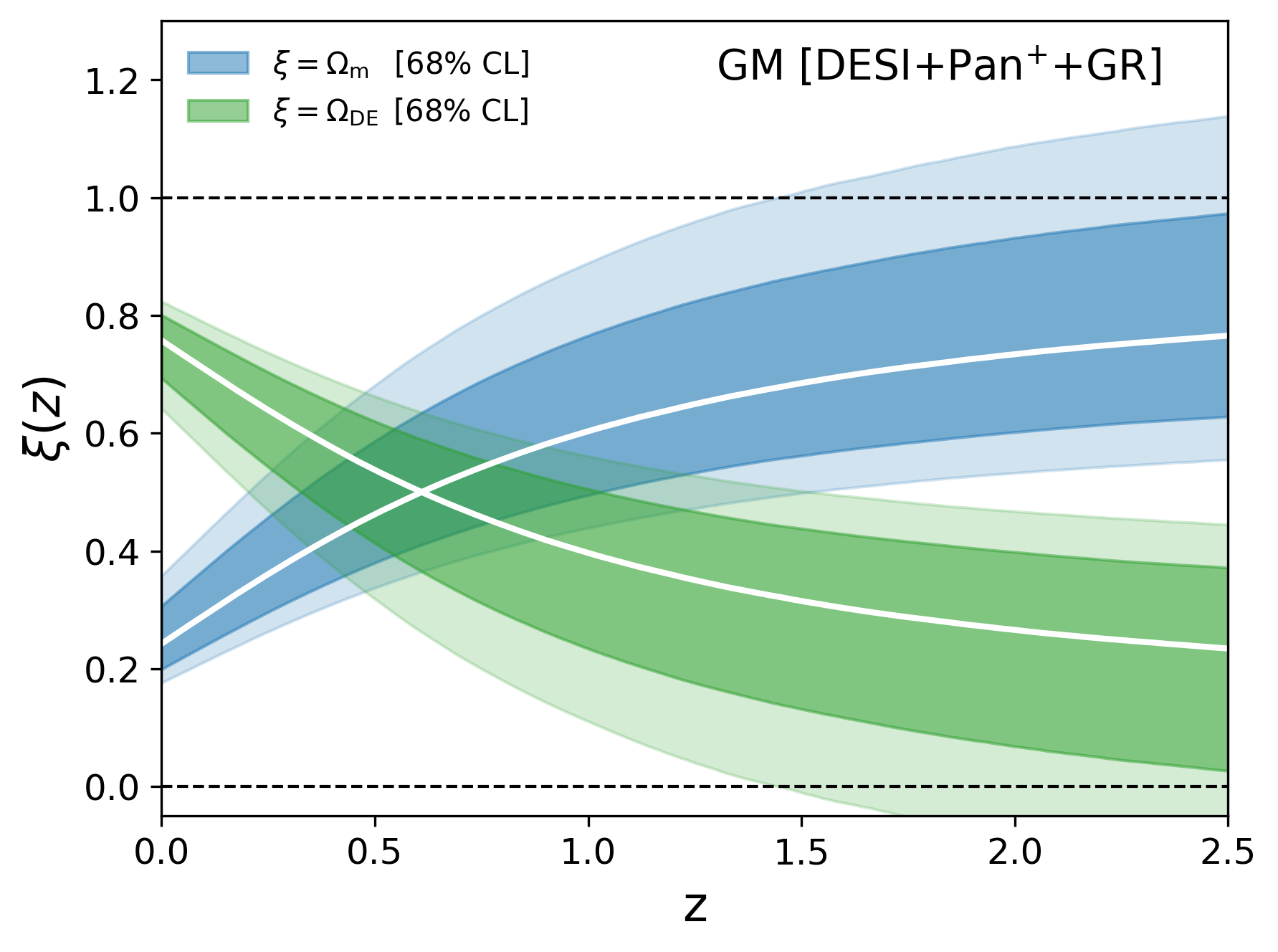}
\includegraphics[scale = 0.38]{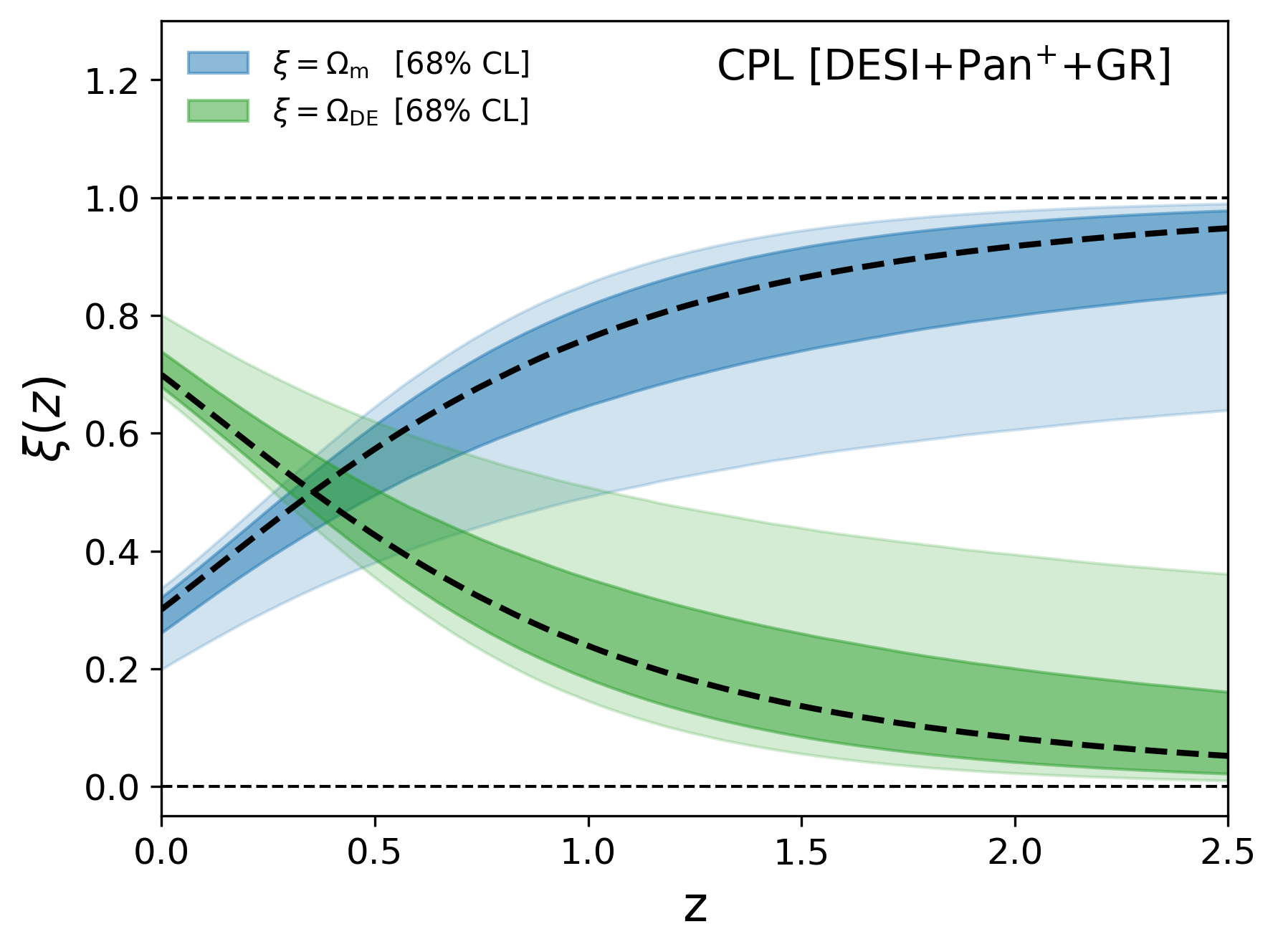}
\includegraphics[scale = 0.38]{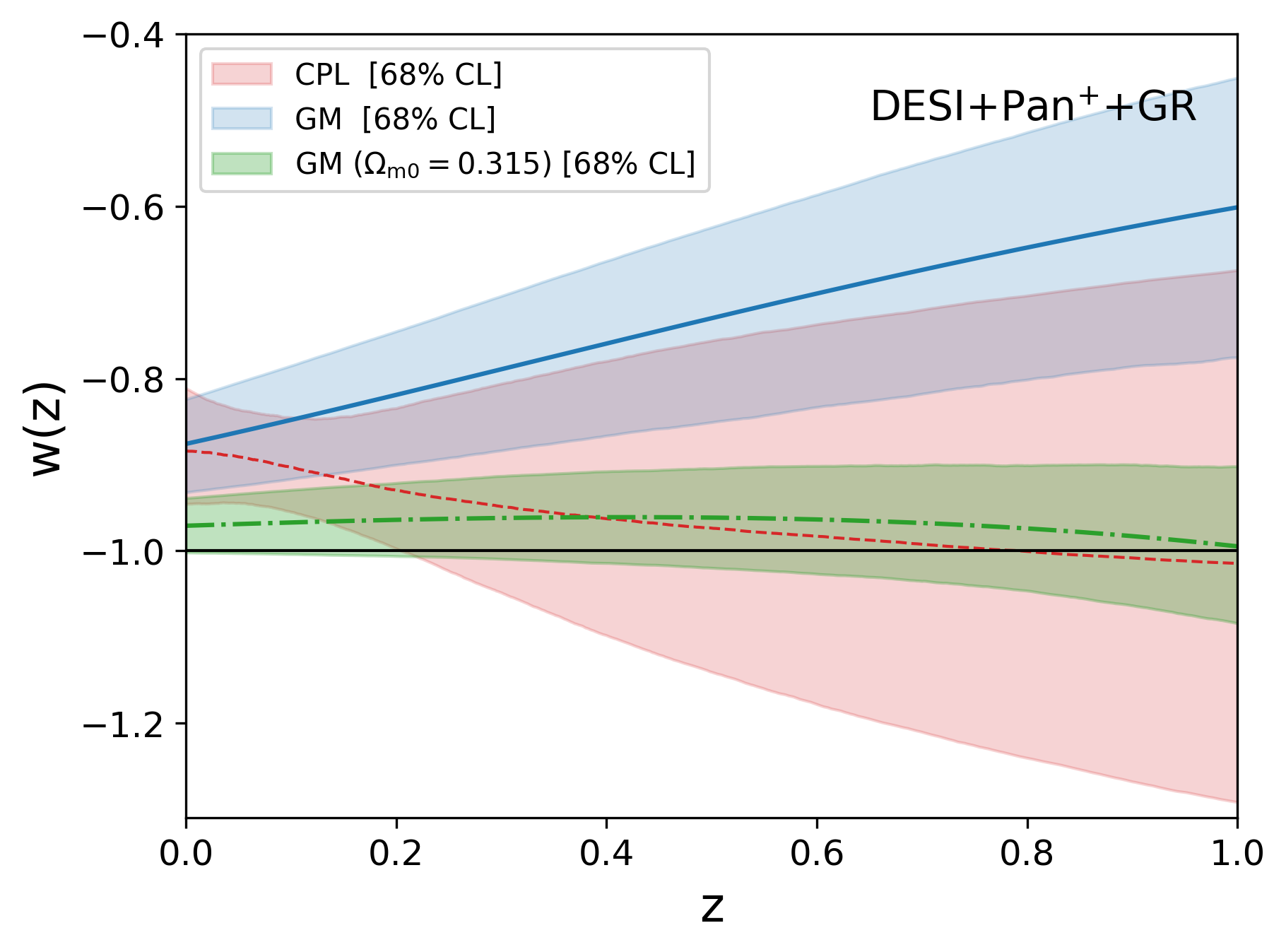}

\caption{\textit{Left}: Variations of dark energy density $\Omega_{\rm DE}$ and 
$\Omega_{\rm m}$ (normalised to the critical density $\rho_{c0}$) with redshift $z$ for DESI+GR+\Panp {for the general model (GM)}. We show the $68\%, 95\%$ C.L. as darker and lighter regions, respectively. {\textit{Center}: Same as left panel but for the CPL model.} \textit{Right}:DE EoS $\omega(z)$ vs redshift $z$ for DESI+GR+\Panp. The shaded region represents $1\sigma$ variation for the CPL parametrization (red) and for our general model (blue) }\label{rhoDE}
\end{figure*}

{This indicates that the earlier SDSS BAO and the well-established \Panp show a mild deviation from the CMB-based \LCDM, while the more recent DESI is more consistent with the standard scenario. Note that this inference is made in our modification of the scale factor in contrast to the evidence for deviations presented in \cite{DESI:2024mwx}, through CPL parametrization.}

In order to extract the DE behaviour (assuming a minimally coupled DE) from the constrained $A-B$ parameter space by BAO+SNIa data, we utilise \cref{EOS} together with the $\Omzero$, which is constrained once the growth rate data is taken into account, as explained above. Prior to studying the redshift evolution, we study how the constrained parameter space $w_{0}-w_{a}$ for a minimally coupled DE compares to the CPL model. To this end, we first constrain the CPL model with DESI+\Panp+Growth rate data and obtain constraints on $\{w_{0},\,w_{a}\}$ and $\Omzero$. We compare the constraints obtained in both the models in the right-top panel of \cref{w0wa_BAOSNGRW}, as one can see, in our model, the correlation\footnote{To test for the effect of the matter density in our general model, on the parameter space $w_{0}-w_{a}$, we also construct the same while assuming the $\Omzero$, constrained using the CPL model, finding only a mild shrinking of the posteriors. } between $w_{0}$ and $w_{a}$ parameters is different than CPL model. Also, the allowed $w_{0}-w_{a}$ space in our general model is much tighter than in the CPL model. Moreover, the major portion of the allowed parameter space now falls in the quintessence region, unlike the CPL case, which indicates a phantom transition. From these, we have two observations: first, the general model has a better constraining power for DE behaviour around the present day compared to CPL model and second, the inferences of the same are extremely dependent on the choice of the parametrization.

{It is even more interesting that DE behaviour at the present day inferred from our general model is now completely in the physically viable region, which was earlier presented in \cite{Raveri:2017qvt}. Note that this did not require any special attention in the model building which uses the simple scale factor parameterization in \cref{eqn:sf_ABCDM}. This in turn is a posterior validation of the model which is capable of providing `necessary' DE dynamics while avoiding the unphysical regions of the parameter space at present day, by construction. To obtain the same within a more flexible CPL parameterization each of the unphysical conditions has to be excluded by hand as shown in \cite{Raveri:2017qvt}. In the right-bottom panel of \cref{w0wa_BAOSNGRW} we show the distribution of parameter $B$ within the posterior of the $w_0- w_{a}$. It is interesting to note that a certain value of parameter $B$, mostly provides an iso-likelihood curve such that the $B= 2/3$ line passes through the \LCDM-case of $\{w_0, w_a\} = \{-1, 0\}$. However, this is only inferred today at $a(t_0) = 1$ and does not imply that the redshift evolution of the DE EoS needs to be completely quintessence-like, as we find that our general model is capable of providing rich phenomenological behaviour. In \cref{Table_bestfit}, we compare the constraints for ($w_{0}$, $w_{a}$, $\Omzero$ and $\sigma_{8}$) between General Model and CPL mode. While the constraints on $w_{0}$ parameter are consistent between the two models, the constraints on $w_{a}$ in these two models for Growth rate+DESI+\Panp combination, can provide varied dynamics in the redshift evolution. Similar to the constraints in the $A-B$ parameter space, we find that the \LCDM case is in agreement only at $\sim 2 \sigma$ level for both DESI and SDSS datasets. Moreover, while the constraints on $\Omzero$ for the general model for two data combinations are in very good agreement, they vary for CPL model. In the last column of the \cref{Table_bestfit}, we also show the Bayesian evidence estimated for both the models and the two dataset combinations, indicating a statistical equivalence. }

{As shown in \cref{rhoDE}, and mentioned earlier it is possible that the $\rhode$ can transition from negative energy density to a positive value. This phenomenon is completely driven by larger values of the $\Omzero$, present in its less precise determination, completely driven by the growth rate data in our analysis. However, it goes on to validate how a late-time dark energy-dominated phenomenon is replicated in our general model. {In the center panel of \cref{rhoDE}, we show the evolution of the matter and dark energy densities within the CPL model, which is completely consistent with evolution in the general model. }{We also note that the departure from the matter-dominated regime occurs slower than in the standard \LCDM or even the CPL model while having the deceleration-to-acceleration transition happening at a redshift $z \sim 0.6$, similar or slightly higher than in the standard scenario, with a larger uncertainty. For instance, from left panel of \cref{rhoDE}, one can infer that within $1\sigma$ the matter-dark energy transition could have occurred around $z\sim 1$. This is completely in accordance with the anticipation as discussed in \cite{Sen_2002}, for larger values of $B>2/3$ when constructing the deceleration parameter. }}

{Finally, in the right panel of \cref{rhoDE} we show the dark energy EoS as inferred from our general model and the CPL model. As already mentioned the the EoS completely depends on the precision inference of the $\Omzero$. We find the DE EoS to be completely consistent with the quintessence in contrast to the CPL-based EoS which transitions to a phantom region around $z\to 1.0$. However, in evaluating the general models' EoS, we have excluded $\sim 15\%$ of the Markovian samples, which present a smaller second mode in the posteriors of $\Omzero$, driven by the growth rate data. {In effect, we filter out the dark energy behaviour that transitions from a $-ve$ to $+ve$ value at a certain redshift ($z_{\pm}$), which if not done can provide singularities in the EoS, emulating several well-known cosmological models such as Brane-world \cite{Sahni:2002dx}. Let us emphasize that in our modelling, this boils down to the precision and our confidence in the constraint of matter density alone.} To further demonstrate the effect of the same, we also evaluate the EoS when assuming a constant $\Omzero = 0.315$, based on Planck-\LCDM value. As anticipated the EoS moves closer to the \LCDM scenario, being completely consistent with $\omega =-1$, within $\sim 1\sigma$. Therefore, as a general behaviour, we infer that the DE EoS moves towards more negative values for larger values of $\Omzero$ once the expansion history itself is constrained by the general model, and even phantom crossing is a possible outcome. Note however that our scale factor parameterization accurately models only the cosmological regime when both dark energy and matter components are of similar order. Keeping this in mind, we conservatively reconstruct the EoS only up to redshift $z<1$, also implying that the inference of the same is more accurate at the present day ($z\to 0$) and low redshifts\footnote{A dedicated investigation of the scale-factor behaviour transitioning from a DE+matter dominated regime to a matter-only epoch is extremely crucial to test the validity of the inferred expansion history at higher redshifts. }. This can also be seen as an excellent agreement between CPL and general models' independent inference of the present-day value of EoS ($w_0$), while they diverge at higher redshifts. }
 
{In \cref{triangle}, we have also shown the constraints on several derived quantities such as $t_{0}H_{0},\, S_8$  and $H_0$ inferred through the inverse distance ladder approach, for the general model. The corresponding numerical constraints are reported in \cref{Table_bestfit}. We find that our general model constraints are mostly in good agreement with the standard Planck-\LCDM scenario. However, particularly in the 2D parameter space of $A-\Hzero$ and $B-\Hzero$, we find a mild $\sim 2 \sigma$ deviation from the standard scenario. We infer that having a better constraint on the $\Omzero$ can provide a clearer implication for the possible deviations, if any. In this context, it is also interesting to note that the growth rate data which is included in our analysis primarily to infer the matter density, does not present any correlations to the parameters $A, B$\footnote{While not immediate one would anticipate a possible positive correlation between the parameters $A$ and $\Omzero$, seeing as to how the parameter $A$ reduces to $\Omzero$ for the standard case.}, which constrain the background evolution.} { We find that our general model, while not capable of entirely alleviating the $\Hzero$-tension \cite{Verde:2023lmm, Riess:2024ohe} or $S_8$-tension \cite{DiValentino:2020vvd}, clearly moves the constraints in the desired direction. However, addressing the tensions is beyond the scope of our motivation here.}

\section{Conclusions}\label{sec:conclusions}
We have presented constraints on the evolution of the scale factor using late-universe probes, namely, SNIa, BAO and growth of structure. Generalising the expression for $a(t)$, we derived the corresponding expression for the evolution of the Hubble parameter with redshift and inferred the parameters defining the scale factor evolution using a combination of BAO and SNIa. We parametrised the evolution of DE density by explicitly introducing a term for the evolution of the matter density with redshift and attributing the rest to the evolution of the expansion history to DE. We found that the DESI BAO + SNe~Ia are consistent with the $A = 0.3, B=2/3$ case, corresponding to the $\Lambda$CDM scenario, at the 1$\sigma$ level. When we use the SDSS BAO instead, the consistency is weaker at the 2$\sigma$ level.
 
From the $\{A,\, B\}$ parameters for the scale factor, we inferred the present-day value of the DE eos for the model and its first-order time derivative. We find that within this model, the degeneracy between the $w_0$ and $w_a$ is orthogonal to the commonly used CPL parametrisation. Moreover, the constraints on the first derivative of the equation of state, $w_a$ are significantly tighter than the CPL parametrisation. We posit that this is due to the assumption of the scale factor evolving as a $sinh$ function of time. This excludes parts of the $w_0$-$w_a$ parameter space that would not correspond to such an evolution of $a(t)$. We find that our constraints also disallow large regions of the $w_0$ - $w_a$ parametrisation that is not physically motivated, e.g. via ghost conditions, the Cauchy problem or exponential growth \citep[e.g., see][]{Peirone2017:Quint}. We, therefore, conclude that more physically motivated parametrisations are necessary to constrain the dark energy equation of state as a function of redshift. In future work, we will explore a larger family of parametrisations for the scale factor and analyse the impact on dark energy inference.

\section*{Acknowledgements} 
\textit{Acknowledgements:} We are grateful to Sunny Vagnozzi and Shahin Sheikh Jabbari for useful discussion and comments on the draft. BSH is supported by the INFN INDARK grant and acknowledges support from the COSMOS project of the Italian Space Agency (cosmosnet.it). SD acknowledges support from a Kavli Fellowship and a Junior Research Fellowship at Lucy Cavendish College. UM and AAS acknowledge the funding from SERB, Govt of India under the research grant no: CRG/2020/004347. AAS also acknowledges support for his visits from INAF - Osservatorio Astronomico di Roma, Rome, Italy, CERN, Switzerland and Abdus Salam International Centre For Theoretical Physics, Trieste, Italy, where part of the work has been done.

\bibliography{reference}

\begin{thebibliography}{61}%
\makeatletter
\providecommand \@ifxundefined [1]{%
 \@ifx{#1\undefined}
}%
\providecommand \@ifnum [1]{%
 \ifnum #1\expandafter \@firstoftwo
 \else \expandafter \@secondoftwo
 \fi
}%
\providecommand \@ifx [1]{%
 \ifx #1\expandafter \@firstoftwo
 \else \expandafter \@secondoftwo
 \fi
}%
\providecommand \natexlab [1]{#1}%
\providecommand \enquote  [1]{``#1''}%
\providecommand \bibnamefont  [1]{#1}%
\providecommand \bibfnamefont [1]{#1}%
\providecommand \citenamefont [1]{#1}%
\providecommand \href@noop [0]{\@secondoftwo}%
\providecommand \href [0]{\begingroup \@sanitize@url \@href}%
\providecommand \@href[1]{\@@startlink{#1}\@@href}%
\providecommand \@@href[1]{\endgroup#1\@@endlink}%
\providecommand \@sanitize@url [0]{\catcode `\\12\catcode `\$12\catcode `\&12\catcode `\#12\catcode `\^12\catcode `\_12\catcode `\%12\relax}%
\providecommand \@@startlink[1]{}%
\providecommand \@@endlink[0]{}%
\providecommand \url  [0]{\begingroup\@sanitize@url \@url }%
\providecommand \@url [1]{\endgroup\@href {#1}{\urlprefix }}%
\providecommand \urlprefix  [0]{URL }%
\providecommand \Eprint [0]{\href }%
\providecommand \doibase [0]{http://dx.doi.org/}%
\providecommand \selectlanguage [0]{\@gobble}%
\providecommand \bibinfo  [0]{\@secondoftwo}%
\providecommand \bibfield  [0]{\@secondoftwo}%
\providecommand \translation [1]{[#1]}%
\providecommand \BibitemOpen [0]{}%
\providecommand \bibitemStop [0]{}%
\providecommand \bibitemNoStop [0]{.\EOS\space}%
\providecommand \EOS [0]{\spacefactor3000\relax}%
\providecommand \BibitemShut  [1]{\csname bibitem#1\endcsname}%
\let\auto@bib@innerbib\@empty
\bibitem [{\citenamefont {{Peebles}}(2024)}]{Peebles2024}%
  \BibitemOpen
  \bibfield  {author} {\bibinfo {author} {\bibfnamefont {P.~J.~E.}\ \bibnamefont {{Peebles}}},\ }\href {\doibase 10.48550/arXiv.2405.18307} {\bibfield  {journal} {\bibinfo  {journal} {arXiv e-prints}\ ,\ \bibinfo {eid} {arXiv:2405.18307}} (\bibinfo {year} {2024})},\ \Eprint {http://arxiv.org/abs/2405.18307} {arXiv:2405.18307 [astro-ph.CO]} \BibitemShut {NoStop}%
\bibitem [{\citenamefont {Riess}\ \emph {et~al.}(1998)\citenamefont {Riess} \emph {et~al.}}]{Riess98}%
  \BibitemOpen
  \bibfield  {author} {\bibinfo {author} {\bibfnamefont {A.~G.}\ \bibnamefont {Riess}} \emph {et~al.} (\bibinfo {collaboration} {Supernova Search Team}),\ }\href {\doibase 10.1086/300499} {\bibfield  {journal} {\bibinfo  {journal} {Astron. J.}\ }\textbf {\bibinfo {volume} {116}},\ \bibinfo {pages} {1009} (\bibinfo {year} {1998})},\ \Eprint {http://arxiv.org/abs/astro-ph/9805201} {arXiv:astro-ph/9805201 [astro-ph]} \BibitemShut {NoStop}%
\bibitem [{\citenamefont {Aghanim}\ \emph {et~al.}(2020)\citenamefont {Aghanim} \emph {et~al.}}]{Planck18_parameters}%
  \BibitemOpen
  \bibfield  {author} {\bibinfo {author} {\bibfnamefont {N.}~\bibnamefont {Aghanim}} \emph {et~al.} (\bibinfo {collaboration} {Planck}),\ }\href {\doibase 10.1051/0004-6361/201833910} {\bibfield  {journal} {\bibinfo  {journal} {Astron. Astrophys.}\ }\textbf {\bibinfo {volume} {641}},\ \bibinfo {pages} {A6} (\bibinfo {year} {2020})},\ \Eprint {http://arxiv.org/abs/1807.06209} {arXiv:1807.06209 [astro-ph.CO]} \BibitemShut {NoStop}%
\bibitem [{\citenamefont {{Planck Collaboration}}\ \emph {et~al.}(2020)\citenamefont {{Planck Collaboration}} \emph {et~al.}}]{Planck2020}%
  \BibitemOpen
  \bibfield  {author} {\bibinfo {author} {\bibnamefont {{Planck Collaboration}}} \emph {et~al.},\ }\href {\doibase 10.1051/0004-6361/201833910} {\bibfield  {journal} {\bibinfo  {journal} {\aap}\ }\textbf {\bibinfo {volume} {641}},\ \bibinfo {eid} {A6} (\bibinfo {year} {2020})},\ \Eprint {http://arxiv.org/abs/1807.06209} {arXiv:1807.06209 [astro-ph.CO]} \BibitemShut {NoStop}%
\bibitem [{\citenamefont {Ade}\ \emph {et~al.}(2016)\citenamefont {Ade} \emph {et~al.}}]{Planck15_DE}%
  \BibitemOpen
  \bibfield  {author} {\bibinfo {author} {\bibfnamefont {P.~A.~R.}\ \bibnamefont {Ade}} \emph {et~al.} (\bibinfo {collaboration} {Planck}),\ }\href {\doibase 10.1051/0004-6361/201525814} {\bibfield  {journal} {\bibinfo  {journal} {Astron. Astrophys.}\ }\textbf {\bibinfo {volume} {594}},\ \bibinfo {pages} {A14} (\bibinfo {year} {2016})},\ \Eprint {http://arxiv.org/abs/1502.01590} {arXiv:1502.01590 [astro-ph.CO]} \BibitemShut {NoStop}%
\bibitem [{\citenamefont {Aiola}\ \emph {et~al.}(2020)\citenamefont {Aiola} \emph {et~al.}}]{ACT:2020gnv}%
  \BibitemOpen
  \bibfield  {author} {\bibinfo {author} {\bibfnamefont {S.}~\bibnamefont {Aiola}} \emph {et~al.} (\bibinfo {collaboration} {ACT}),\ }\href {\doibase 10.1088/1475-7516/2020/12/047} {\bibfield  {journal} {\bibinfo  {journal} {JCAP}\ }\textbf {\bibinfo {volume} {12}},\ \bibinfo {pages} {047} (\bibinfo {year} {2020})},\ \Eprint {http://arxiv.org/abs/2007.07288} {arXiv:2007.07288 [astro-ph.CO]} \BibitemShut {NoStop}%
\bibitem [{\citenamefont {Aubourg}\ \emph {et~al.}(2015)\citenamefont {Aubourg}, \citenamefont {Bailey}, \citenamefont {Bautista}, \citenamefont {Beutler}, \citenamefont {Bhardwaj}, \citenamefont {Bizyaev}, \citenamefont {Blanton}, \citenamefont {Blomqvist}, \citenamefont {Bolton}, \citenamefont {Bovy},\ and\ \citenamefont {et~al.}}]{Aubourg_2015}%
  \BibitemOpen
  \bibfield  {author} {\bibinfo {author} {\bibfnamefont {{\'E}.}~\bibnamefont {Aubourg}}, \bibinfo {author} {\bibfnamefont {S.}~\bibnamefont {Bailey}}, \bibinfo {author} {\bibfnamefont {J.~E.}\ \bibnamefont {Bautista}}, \bibinfo {author} {\bibfnamefont {F.}~\bibnamefont {Beutler}}, \bibinfo {author} {\bibfnamefont {V.}~\bibnamefont {Bhardwaj}}, \bibinfo {author} {\bibfnamefont {D.}~\bibnamefont {Bizyaev}}, \bibinfo {author} {\bibfnamefont {M.}~\bibnamefont {Blanton}}, \bibinfo {author} {\bibfnamefont {M.}~\bibnamefont {Blomqvist}}, \bibinfo {author} {\bibfnamefont {A.~S.}\ \bibnamefont {Bolton}}, \bibinfo {author} {\bibfnamefont {J.}~\bibnamefont {Bovy}}, \ and\ \bibinfo {author} {\bibnamefont {et~al.}},\ }\href {\doibase 10.1103/physrevd.92.123516} {\bibfield  {journal} {\bibinfo  {journal} {Physical Review D}\ }\textbf {\bibinfo {volume} {92}} (\bibinfo {year} {2015}),\ 10.1103/physrevd.92.123516}\BibitemShut {NoStop}%
\bibitem [{\citenamefont {Alam}\ \emph {et~al.}(2017)\citenamefont {Alam} \emph {et~al.}}]{Alam16}%
  \BibitemOpen
  \bibfield  {author} {\bibinfo {author} {\bibfnamefont {S.}~\bibnamefont {Alam}} \emph {et~al.} (\bibinfo {collaboration} {BOSS}),\ }\href {\doibase 10.1093/mnras/stx721} {\bibfield  {journal} {\bibinfo  {journal} {Mon. Not. Roy. Astron. Soc.}\ }\textbf {\bibinfo {volume} {470}},\ \bibinfo {pages} {2617} (\bibinfo {year} {2017})},\ \Eprint {http://arxiv.org/abs/1607.03155} {arXiv:1607.03155 [astro-ph.CO]} \BibitemShut {NoStop}%
\bibitem [{\citenamefont {Addison}\ \emph {et~al.}(2018)\citenamefont {Addison}, \citenamefont {Watts}, \citenamefont {Bennett}, \citenamefont {Halpern}, \citenamefont {Hinshaw},\ and\ \citenamefont {Weiland}}]{Addison:2017fdm}%
  \BibitemOpen
  \bibfield  {author} {\bibinfo {author} {\bibfnamefont {G.~E.}\ \bibnamefont {Addison}}, \bibinfo {author} {\bibfnamefont {D.~J.}\ \bibnamefont {Watts}}, \bibinfo {author} {\bibfnamefont {C.~L.}\ \bibnamefont {Bennett}}, \bibinfo {author} {\bibfnamefont {M.}~\bibnamefont {Halpern}}, \bibinfo {author} {\bibfnamefont {G.}~\bibnamefont {Hinshaw}}, \ and\ \bibinfo {author} {\bibfnamefont {J.~L.}\ \bibnamefont {Weiland}},\ }\href {\doibase 10.3847/1538-4357/aaa1ed} {\bibfield  {journal} {\bibinfo  {journal} {Astrophys. J.}\ }\textbf {\bibinfo {volume} {853}},\ \bibinfo {pages} {119} (\bibinfo {year} {2018})},\ \Eprint {http://arxiv.org/abs/1707.06547} {arXiv:1707.06547 [astro-ph.CO]} \BibitemShut {NoStop}%
\bibitem [{\citenamefont {Haridasu}\ \emph {et~al.}(2018)\citenamefont {Haridasu}, \citenamefont {Lukovi{\'{c}}},\ and\ \citenamefont {Vittorio}}]{Haridasu17_bao}%
  \BibitemOpen
  \bibfield  {author} {\bibinfo {author} {\bibfnamefont {B.~S.}\ \bibnamefont {Haridasu}}, \bibinfo {author} {\bibfnamefont {V.~V.}\ \bibnamefont {Lukovi{\'{c}}}}, \ and\ \bibinfo {author} {\bibfnamefont {N.}~\bibnamefont {Vittorio}},\ }\href {\doibase 10.1088/1475-7516/2018/05/033} {\bibfield  {journal} {\bibinfo  {journal} {JCAP}\ }\textbf {\bibinfo {volume} {1805}},\ \bibinfo {pages} {033} (\bibinfo {year} {2018})},\ \Eprint {http://arxiv.org/abs/1711.03929} {arXiv:1711.03929 [astro-ph.CO]} \BibitemShut {NoStop}%
\bibitem [{\citenamefont {Riess}(2019)}]{Riess:2019qba}%
  \BibitemOpen
  \bibfield  {author} {\bibinfo {author} {\bibfnamefont {A.~G.}\ \bibnamefont {Riess}},\ }\href {\doibase 10.1038/s42254-019-0137-0} {\bibfield  {journal} {\bibinfo  {journal} {Nature Rev. Phys.}\ }\textbf {\bibinfo {volume} {2}},\ \bibinfo {pages} {10} (\bibinfo {year} {2019})},\ \Eprint {http://arxiv.org/abs/2001.03624} {arXiv:2001.03624 [astro-ph.CO]} \BibitemShut {NoStop}%
\bibitem [{\citenamefont {Scolnic}\ \emph {et~al.}(2018)\citenamefont {Scolnic} \emph {et~al.}}]{Scolnic:2017caz}%
  \BibitemOpen
  \bibfield  {author} {\bibinfo {author} {\bibfnamefont {D.~M.}\ \bibnamefont {Scolnic}} \emph {et~al.},\ }\href {\doibase 10.3847/1538-4357/aab9bb} {\bibfield  {journal} {\bibinfo  {journal} {Astrophys. J.}\ }\textbf {\bibinfo {volume} {859}},\ \bibinfo {pages} {101} (\bibinfo {year} {2018})},\ \Eprint {http://arxiv.org/abs/1710.00845} {arXiv:1710.00845 [astro-ph.CO]} \BibitemShut {NoStop}%
\bibitem [{\citenamefont {Brout}\ \emph {et~al.}(2022)\citenamefont {Brout} \emph {et~al.}}]{Brout:2022vxf}%
  \BibitemOpen
  \bibfield  {author} {\bibinfo {author} {\bibfnamefont {D.}~\bibnamefont {Brout}} \emph {et~al.},\ }\href {\doibase 10.3847/1538-4357/ac8e04} {\bibfield  {journal} {\bibinfo  {journal} {Astrophys. J.}\ }\textbf {\bibinfo {volume} {938}},\ \bibinfo {pages} {110} (\bibinfo {year} {2022})},\ \Eprint {http://arxiv.org/abs/2202.04077} {arXiv:2202.04077 [astro-ph.CO]} \BibitemShut {NoStop}%
\bibitem [{\citenamefont {Adame}\ \emph {et~al.}(2024)\citenamefont {Adame} \emph {et~al.}}]{DESI:2024mwx}%
  \BibitemOpen
  \bibfield  {author} {\bibinfo {author} {\bibfnamefont {A.~G.}\ \bibnamefont {Adame}} \emph {et~al.} (\bibinfo {collaboration} {DESI}),\ }\href@noop {} {\  (\bibinfo {year} {2024})},\ \Eprint {http://arxiv.org/abs/2404.03002} {arXiv:2404.03002 [astro-ph.CO]} \BibitemShut {NoStop}%
\bibitem [{\citenamefont {Rubin}\ \emph {et~al.}(2023)\citenamefont {Rubin} \emph {et~al.}}]{Rubin:2023ovl}%
  \BibitemOpen
  \bibfield  {author} {\bibinfo {author} {\bibfnamefont {D.}~\bibnamefont {Rubin}} \emph {et~al.},\ }\href@noop {} {\  (\bibinfo {year} {2023})},\ \Eprint {http://arxiv.org/abs/2311.12098} {arXiv:2311.12098 [astro-ph.CO]} \BibitemShut {NoStop}%
\bibitem [{\citenamefont {Abbott}\ \emph {et~al.}(2024)\citenamefont {Abbott} \emph {et~al.}}]{DES:2024tys}%
  \BibitemOpen
  \bibfield  {author} {\bibinfo {author} {\bibfnamefont {T.~M.~C.}\ \bibnamefont {Abbott}} \emph {et~al.} (\bibinfo {collaboration} {DES}),\ }\href@noop {} {\  (\bibinfo {year} {2024})},\ \Eprint {http://arxiv.org/abs/2401.02929} {arXiv:2401.02929 [astro-ph.CO]} \BibitemShut {NoStop}%
\bibitem [{\citenamefont {{Riess}}\ \emph {et~al.}(2022)\citenamefont {{Riess}} \emph {et~al.}}]{Riess2022}%
  \BibitemOpen
  \bibfield  {author} {\bibinfo {author} {\bibfnamefont {A.~G.}\ \bibnamefont {{Riess}}} \emph {et~al.},\ }\href {\doibase 10.3847/2041-8213/ac5c5b10.48550/arXiv.2112.04510} {\bibfield  {journal} {\bibinfo  {journal} {\apjl}\ }\textbf {\bibinfo {volume} {934}},\ \bibinfo {eid} {L7} (\bibinfo {year} {2022})},\ \Eprint {http://arxiv.org/abs/2112.04510} {arXiv:2112.04510 [astro-ph.CO]} \BibitemShut {NoStop}%
\bibitem [{\citenamefont {{Weinberg}}(2000)}]{Weinberg2000}%
  \BibitemOpen
  \bibfield  {author} {\bibinfo {author} {\bibfnamefont {S.}~\bibnamefont {{Weinberg}}},\ }\href {\doibase 10.48550/arXiv.astro-ph/0005265} {\bibfield  {journal} {\bibinfo  {journal} {arXiv e-prints}\ ,\ \bibinfo {eid} {astro-ph/0005265}} (\bibinfo {year} {2000})},\ \Eprint {http://arxiv.org/abs/astro-ph/0005265} {arXiv:astro-ph/0005265 [astro-ph]} \BibitemShut {NoStop}%
\bibitem [{\citenamefont {CHEVALLIER}\ and\ \citenamefont {POLARSKI}(2001)}]{CHEVALLIER_2001}%
  \BibitemOpen
  \bibfield  {author} {\bibinfo {author} {\bibfnamefont {M.}~\bibnamefont {CHEVALLIER}}\ and\ \bibinfo {author} {\bibfnamefont {D.}~\bibnamefont {POLARSKI}},\ }\href {\doibase 10.1142/s0218271801000822} {\bibfield  {journal} {\bibinfo  {journal} {International Journal of Modern Physics D}\ }\textbf {\bibinfo {volume} {10}},\ \bibinfo {pages} {213–223} (\bibinfo {year} {2001})}\BibitemShut {NoStop}%
\bibitem [{\citenamefont {Linder}(2003)}]{Linder_2003}%
  \BibitemOpen
  \bibfield  {author} {\bibinfo {author} {\bibfnamefont {E.~V.}\ \bibnamefont {Linder}},\ }\href {\doibase 10.1103/physrevlett.90.091301} {\bibfield  {journal} {\bibinfo  {journal} {Physical Review Letters}\ }\textbf {\bibinfo {volume} {90}} (\bibinfo {year} {2003}),\ 10.1103/physrevlett.90.091301}\BibitemShut {NoStop}%
\bibitem [{\citenamefont {Linder}(2005)}]{Linder:2005in}%
  \BibitemOpen
  \bibfield  {author} {\bibinfo {author} {\bibfnamefont {E.~V.}\ \bibnamefont {Linder}},\ }\href {\doibase 10.1103/PhysRevD.72.043529} {\bibfield  {journal} {\bibinfo  {journal} {Phys. Rev. D}\ }\textbf {\bibinfo {volume} {72}},\ \bibinfo {pages} {043529} (\bibinfo {year} {2005})},\ \Eprint {http://arxiv.org/abs/astro-ph/0507263} {arXiv:astro-ph/0507263} \BibitemShut {NoStop}%
\bibitem [{\citenamefont {{Linder}}(2003)}]{Linder02}%
  \BibitemOpen
  \bibfield  {author} {\bibinfo {author} {\bibfnamefont {E.~V.}\ \bibnamefont {{Linder}}},\ }\href {\doibase 10.1103/PhysRevLett.90.091301} {\bibfield  {journal} {\bibinfo  {journal} {Physical Review Letters}\ }\textbf {\bibinfo {volume} {90}},\ \bibinfo {eid} {091301} (\bibinfo {year} {2003})},\ \Eprint {http://arxiv.org/abs/astro-ph/0208512} {astro-ph/0208512} \BibitemShut {NoStop}%
\bibitem [{\citenamefont {Capozziello}\ \emph {et~al.}(2006)\citenamefont {Capozziello}, \citenamefont {Cardone}, \citenamefont {Piedipalumbo},\ and\ \citenamefont {Rubano}}]{Capozziello06}%
  \BibitemOpen
  \bibfield  {author} {\bibinfo {author} {\bibfnamefont {S.}~\bibnamefont {Capozziello}}, \bibinfo {author} {\bibfnamefont {V.~F.}\ \bibnamefont {Cardone}}, \bibinfo {author} {\bibfnamefont {E.}~\bibnamefont {Piedipalumbo}}, \ and\ \bibinfo {author} {\bibfnamefont {C.}~\bibnamefont {Rubano}},\ }\href {\doibase 10.1088/0264-9381/23/4/009} {\bibfield  {journal} {\bibinfo  {journal} {Classical and Quantum Gravity}\ }\textbf {\bibinfo {volume} {23}},\ \bibinfo {pages} {1205} (\bibinfo {year} {2006})},\ \Eprint {http://arxiv.org/abs/astro-ph/0507438} {astro-ph/0507438} \BibitemShut {NoStop}%
\bibitem [{\citenamefont {Sotiriou}\ and\ \citenamefont {Faraoni}(2010)}]{Sotiriou_2010}%
  \BibitemOpen
  \bibfield  {author} {\bibinfo {author} {\bibfnamefont {T.~P.}\ \bibnamefont {Sotiriou}}\ and\ \bibinfo {author} {\bibfnamefont {V.}~\bibnamefont {Faraoni}},\ }\href {\doibase 10.1103/revmodphys.82.451} {\bibfield  {journal} {\bibinfo  {journal} {Reviews of Modern Physics}\ }\textbf {\bibinfo {volume} {82}},\ \bibinfo {pages} {451–497} (\bibinfo {year} {2010})}\BibitemShut {NoStop}%
\bibitem [{\citenamefont {Nojiri}\ \emph {et~al.}(2017)\citenamefont {Nojiri}, \citenamefont {Odintsov},\ and\ \citenamefont {Oikonomou}}]{Nojiri:2017ncd}%
  \BibitemOpen
  \bibfield  {author} {\bibinfo {author} {\bibfnamefont {S.}~\bibnamefont {Nojiri}}, \bibinfo {author} {\bibfnamefont {S.~D.}\ \bibnamefont {Odintsov}}, \ and\ \bibinfo {author} {\bibfnamefont {V.~K.}\ \bibnamefont {Oikonomou}},\ }\href {\doibase 10.1016/j.physrep.2017.06.001} {\bibfield  {journal} {\bibinfo  {journal} {Phys. Rept.}\ }\textbf {\bibinfo {volume} {692}},\ \bibinfo {pages} {1} (\bibinfo {year} {2017})},\ \Eprint {http://arxiv.org/abs/1705.11098} {arXiv:1705.11098 [gr-qc]} \BibitemShut {NoStop}%
\bibitem [{\citenamefont {Yasunori}\ and\ \citenamefont {Kei-ichi}(2003)}]{YasunoriFujii_2003}%
  \BibitemOpen
  \bibfield  {author} {\bibinfo {author} {\bibfnamefont {F.}~\bibnamefont {Yasunori}}\ and\ \bibinfo {author} {\bibfnamefont {M.}~\bibnamefont {Kei-ichi}},\ }\href {\doibase 10.1088/0264-9381/20/20/601} {\bibfield  {journal} {\bibinfo  {journal} {Classical and Quantum Gravity}\ }\textbf {\bibinfo {volume} {20}},\ \bibinfo {pages} {4503} (\bibinfo {year} {2003})}\BibitemShut {NoStop}%
\bibitem [{\citenamefont {Freese}\ and\ \citenamefont {Lewis}(2002)}]{Freese_2002}%
  \BibitemOpen
  \bibfield  {author} {\bibinfo {author} {\bibfnamefont {K.}~\bibnamefont {Freese}}\ and\ \bibinfo {author} {\bibfnamefont {M.}~\bibnamefont {Lewis}},\ }\href {\doibase 10.1016/s0370-2693(02)02122-6} {\bibfield  {journal} {\bibinfo  {journal} {Physics Letters B}\ }\textbf {\bibinfo {volume} {540}},\ \bibinfo {pages} {1–8} (\bibinfo {year} {2002})}\BibitemShut {NoStop}%
\bibitem [{\citenamefont {Dvali}\ \emph {et~al.}(2000)\citenamefont {Dvali}, \citenamefont {Gabadadze},\ and\ \citenamefont {Porrati}}]{Dvali:2000hr}%
  \BibitemOpen
  \bibfield  {author} {\bibinfo {author} {\bibfnamefont {G.~R.}\ \bibnamefont {Dvali}}, \bibinfo {author} {\bibfnamefont {G.}~\bibnamefont {Gabadadze}}, \ and\ \bibinfo {author} {\bibfnamefont {M.}~\bibnamefont {Porrati}},\ }\href {\doibase 10.1016/S0370-2693(00)00669-9} {\bibfield  {journal} {\bibinfo  {journal} {Phys. Lett. B}\ }\textbf {\bibinfo {volume} {485}},\ \bibinfo {pages} {208} (\bibinfo {year} {2000})},\ \Eprint {http://arxiv.org/abs/hep-th/0005016} {arXiv:hep-th/0005016} \BibitemShut {NoStop}%
\bibitem [{\citenamefont {Barreiro}\ and\ \citenamefont {Sen}(2004)}]{Barreiro_2004}%
  \BibitemOpen
  \bibfield  {author} {\bibinfo {author} {\bibfnamefont {T.}~\bibnamefont {Barreiro}}\ and\ \bibinfo {author} {\bibfnamefont {A.~A.}\ \bibnamefont {Sen}},\ }\href {\doibase 10.1103/physrevd.70.124013} {\bibfield  {journal} {\bibinfo  {journal} {Physical Review D}\ }\textbf {\bibinfo {volume} {70}} (\bibinfo {year} {2004}),\ 10.1103/physrevd.70.124013}\BibitemShut {NoStop}%
\bibitem [{\citenamefont {Vattis}\ \emph {et~al.}(2019)\citenamefont {Vattis}, \citenamefont {Koushiappas},\ and\ \citenamefont {Loeb}}]{Vattis:2019efj}%
  \BibitemOpen
  \bibfield  {author} {\bibinfo {author} {\bibfnamefont {K.}~\bibnamefont {Vattis}}, \bibinfo {author} {\bibfnamefont {S.~M.}\ \bibnamefont {Koushiappas}}, \ and\ \bibinfo {author} {\bibfnamefont {A.}~\bibnamefont {Loeb}},\ }\href {\doibase 10.1103/PhysRevD.99.121302} {\bibfield  {journal} {\bibinfo  {journal} {Phys. Rev. D}\ }\textbf {\bibinfo {volume} {99}},\ \bibinfo {pages} {121302} (\bibinfo {year} {2019})},\ \Eprint {http://arxiv.org/abs/1903.06220} {arXiv:1903.06220 [astro-ph.CO]} \BibitemShut {NoStop}%
\bibitem [{\citenamefont {Clark}\ \emph {et~al.}(2020)\citenamefont {Clark}, \citenamefont {Vattis},\ and\ \citenamefont {Koushiappas}}]{Clark20}%
  \BibitemOpen
  \bibfield  {author} {\bibinfo {author} {\bibfnamefont {S.~J.}\ \bibnamefont {Clark}}, \bibinfo {author} {\bibfnamefont {K.}~\bibnamefont {Vattis}}, \ and\ \bibinfo {author} {\bibfnamefont {S.~M.}\ \bibnamefont {Koushiappas}},\ }\href@noop {} {\  (\bibinfo {year} {2020})},\ \Eprint {http://arxiv.org/abs/2006.03678} {arXiv:2006.03678 [astro-ph.CO]} \BibitemShut {NoStop}%
\bibitem [{\citenamefont {Haridasu}\ and\ \citenamefont {Viel}(2020)}]{Haridasu:2020xaa}%
  \BibitemOpen
  \bibfield  {author} {\bibinfo {author} {\bibfnamefont {B.~S.}\ \bibnamefont {Haridasu}}\ and\ \bibinfo {author} {\bibfnamefont {M.}~\bibnamefont {Viel}},\ }\href {\doibase 10.1093/mnras/staa1991} {\bibfield  {journal} {\bibinfo  {journal} {Mon. Not. Roy. Astron. Soc.}\ }\textbf {\bibinfo {volume} {497}},\ \bibinfo {pages} {1757} (\bibinfo {year} {2020})},\ \Eprint {http://arxiv.org/abs/2004.07709} {arXiv:2004.07709 [astro-ph.CO]} \BibitemShut {NoStop}%
\bibitem [{\citenamefont {Franco~Abell\'an}\ \emph {et~al.}(2022)\citenamefont {Franco~Abell\'an}, \citenamefont {Murgia}, \citenamefont {Poulin},\ and\ \citenamefont {Lavalle}}]{FrancoAbellan:2020xnr}%
  \BibitemOpen
  \bibfield  {author} {\bibinfo {author} {\bibfnamefont {G.}~\bibnamefont {Franco~Abell\'an}}, \bibinfo {author} {\bibfnamefont {R.}~\bibnamefont {Murgia}}, \bibinfo {author} {\bibfnamefont {V.}~\bibnamefont {Poulin}}, \ and\ \bibinfo {author} {\bibfnamefont {J.}~\bibnamefont {Lavalle}},\ }\href {\doibase 10.1103/PhysRevD.105.063525} {\bibfield  {journal} {\bibinfo  {journal} {Phys. Rev. D}\ }\textbf {\bibinfo {volume} {105}},\ \bibinfo {pages} {063525} (\bibinfo {year} {2022})},\ \Eprint {http://arxiv.org/abs/2008.09615} {arXiv:2008.09615 [astro-ph.CO]} \BibitemShut {NoStop}%
\bibitem [{\citenamefont {Naidoo}\ \emph {et~al.}(2024)\citenamefont {Naidoo}, \citenamefont {Jaber}, \citenamefont {Hellwing},\ and\ \citenamefont {Bilicki}}]{Naidoo:2022rda}%
  \BibitemOpen
  \bibfield  {author} {\bibinfo {author} {\bibfnamefont {K.}~\bibnamefont {Naidoo}}, \bibinfo {author} {\bibfnamefont {M.}~\bibnamefont {Jaber}}, \bibinfo {author} {\bibfnamefont {W.~A.}\ \bibnamefont {Hellwing}}, \ and\ \bibinfo {author} {\bibfnamefont {M.}~\bibnamefont {Bilicki}},\ }\href {\doibase 10.1103/PhysRevD.109.083511} {\bibfield  {journal} {\bibinfo  {journal} {Phys. Rev. D}\ }\textbf {\bibinfo {volume} {109}},\ \bibinfo {pages} {083511} (\bibinfo {year} {2024})},\ \Eprint {http://arxiv.org/abs/2209.08102} {arXiv:2209.08102 [astro-ph.CO]} \BibitemShut {NoStop}%
\bibitem [{\citenamefont {Poulot}\ \emph {et~al.}(2024)\citenamefont {Poulot}, \citenamefont {Teixeira}, \citenamefont {van~de Bruck},\ and\ \citenamefont {Nunes}}]{Poulot:2024sex}%
  \BibitemOpen
  \bibfield  {author} {\bibinfo {author} {\bibfnamefont {G.}~\bibnamefont {Poulot}}, \bibinfo {author} {\bibfnamefont {E.~M.}\ \bibnamefont {Teixeira}}, \bibinfo {author} {\bibfnamefont {C.}~\bibnamefont {van~de Bruck}}, \ and\ \bibinfo {author} {\bibfnamefont {N.~J.}\ \bibnamefont {Nunes}},\ }\href@noop {} {\  (\bibinfo {year} {2024})},\ \Eprint {http://arxiv.org/abs/2404.10524} {arXiv:2404.10524 [astro-ph.CO]} \BibitemShut {NoStop}%
\bibitem [{\citenamefont {Lapi}\ \emph {et~al.}(2023)\citenamefont {Lapi}, \citenamefont {Boco}, \citenamefont {Cueli}, \citenamefont {Haridasu}, \citenamefont {Ronconi}, \citenamefont {Baccigalupi},\ and\ \citenamefont {Danese}}]{Lapi:2023plb}%
  \BibitemOpen
  \bibfield  {author} {\bibinfo {author} {\bibfnamefont {A.}~\bibnamefont {Lapi}}, \bibinfo {author} {\bibfnamefont {L.}~\bibnamefont {Boco}}, \bibinfo {author} {\bibfnamefont {M.~M.}\ \bibnamefont {Cueli}}, \bibinfo {author} {\bibfnamefont {B.~S.}\ \bibnamefont {Haridasu}}, \bibinfo {author} {\bibfnamefont {T.}~\bibnamefont {Ronconi}}, \bibinfo {author} {\bibfnamefont {C.}~\bibnamefont {Baccigalupi}}, \ and\ \bibinfo {author} {\bibfnamefont {L.}~\bibnamefont {Danese}},\ }\href {\doibase 10.3847/1538-4357/ad01bb} {\bibfield  {journal} {\bibinfo  {journal} {Astrophys. J.}\ }\textbf {\bibinfo {volume} {959}},\ \bibinfo {pages} {83} (\bibinfo {year} {2023})},\ \Eprint {http://arxiv.org/abs/2310.06028} {arXiv:2310.06028 [astro-ph.CO]} \BibitemShut {NoStop}%
\bibitem [{\citenamefont {Colg\'ain}\ \emph {et~al.}(2023)\citenamefont {Colg\'ain}, \citenamefont {Sheikh-Jabbari},\ and\ \citenamefont {Solomon}}]{Colgain:2022tql}%
  \BibitemOpen
  \bibfield  {author} {\bibinfo {author} {\bibfnamefont {E.~O.}\ \bibnamefont {Colg\'ain}}, \bibinfo {author} {\bibfnamefont {M.~M.}\ \bibnamefont {Sheikh-Jabbari}}, \ and\ \bibinfo {author} {\bibfnamefont {R.}~\bibnamefont {Solomon}},\ }\href {\doibase 10.1016/j.dark.2023.101216} {\bibfield  {journal} {\bibinfo  {journal} {Phys. Dark Univ.}\ }\textbf {\bibinfo {volume} {40}},\ \bibinfo {pages} {101216} (\bibinfo {year} {2023})},\ \Eprint {http://arxiv.org/abs/2211.02129} {arXiv:2211.02129 [astro-ph.CO]} \BibitemShut {NoStop}%
\bibitem [{\citenamefont {Zhumabek}\ \emph {et~al.}(2024)\citenamefont {Zhumabek}, \citenamefont {Denissenya},\ and\ \citenamefont {Linder}}]{Zhumabek:2023ejd}%
  \BibitemOpen
  \bibfield  {author} {\bibinfo {author} {\bibfnamefont {T.}~\bibnamefont {Zhumabek}}, \bibinfo {author} {\bibfnamefont {M.}~\bibnamefont {Denissenya}}, \ and\ \bibinfo {author} {\bibfnamefont {E.~V.}\ \bibnamefont {Linder}},\ }\href {\doibase 10.1088/1475-7516/2024/02/018} {\bibfield  {journal} {\bibinfo  {journal} {JCAP}\ }\textbf {\bibinfo {volume} {02}},\ \bibinfo {pages} {018} (\bibinfo {year} {2024})},\ \Eprint {http://arxiv.org/abs/2311.13795} {arXiv:2311.13795 [astro-ph.CO]} \BibitemShut {NoStop}%
\bibitem [{\citenamefont {Reid}\ \emph {et~al.}(2002)\citenamefont {Reid}, \citenamefont {Kittell}, \citenamefont {Arsznov},\ and\ \citenamefont {Thompson}}]{Reid:2002kp}%
  \BibitemOpen
  \bibfield  {author} {\bibinfo {author} {\bibfnamefont {D.~D.}\ \bibnamefont {Reid}}, \bibinfo {author} {\bibfnamefont {D.~W.}\ \bibnamefont {Kittell}}, \bibinfo {author} {\bibfnamefont {E.~E.}\ \bibnamefont {Arsznov}}, \ and\ \bibinfo {author} {\bibfnamefont {G.~B.}\ \bibnamefont {Thompson}},\ }\href@noop {} {\  (\bibinfo {year} {2002})},\ \Eprint {http://arxiv.org/abs/astro-ph/0209504} {arXiv:astro-ph/0209504} \BibitemShut {NoStop}%
\bibitem [{\citenamefont {Gron}(2002)}]{Gron:2002wrd}%
  \BibitemOpen
  \bibfield  {author} {\bibinfo {author} {\bibfnamefont {O.}~\bibnamefont {Gron}},\ }\href {\doibase 10.1088/0143-0807/23/2/307} {\bibfield  {journal} {\bibinfo  {journal} {Eur. J. Phys.}\ }\textbf {\bibinfo {volume} {23}},\ \bibinfo {pages} {135} (\bibinfo {year} {2002})},\ \Eprint {http://arxiv.org/abs/0801.0552} {arXiv:0801.0552 [astro-ph]} \BibitemShut {NoStop}%
\bibitem [{\citenamefont {Anselmi}\ \emph {et~al.}(2023)\citenamefont {Anselmi}, \citenamefont {Carney}, \citenamefont {Giblin}, \citenamefont {Kumar}, \citenamefont {Mertens}, \citenamefont {O'Dwyer}, \citenamefont {Starkman},\ and\ \citenamefont {Tian}}]{Anselmi:2022uvj}%
  \BibitemOpen
  \bibfield  {author} {\bibinfo {author} {\bibfnamefont {S.}~\bibnamefont {Anselmi}}, \bibinfo {author} {\bibfnamefont {M.~F.}\ \bibnamefont {Carney}}, \bibinfo {author} {\bibfnamefont {J.~T.}\ \bibnamefont {Giblin}}, \bibinfo {author} {\bibfnamefont {S.}~\bibnamefont {Kumar}}, \bibinfo {author} {\bibfnamefont {J.~B.}\ \bibnamefont {Mertens}}, \bibinfo {author} {\bibfnamefont {M.}~\bibnamefont {O'Dwyer}}, \bibinfo {author} {\bibfnamefont {G.~D.}\ \bibnamefont {Starkman}}, \ and\ \bibinfo {author} {\bibfnamefont {C.}~\bibnamefont {Tian}},\ }\href {\doibase 10.1088/1475-7516/2023/02/049} {\bibfield  {journal} {\bibinfo  {journal} {JCAP}\ }\textbf {\bibinfo {volume} {02}},\ \bibinfo {pages} {049} (\bibinfo {year} {2023})},\ \Eprint {http://arxiv.org/abs/2207.06547} {arXiv:2207.06547 [astro-ph.CO]} \BibitemShut {NoStop}%
\bibitem [{\citenamefont {Jimenez}\ \emph {et~al.}(2023)\citenamefont {Jimenez}, \citenamefont {Khalifeh}, \citenamefont {Litim}, \citenamefont {Matarrese},\ and\ \citenamefont {Wandelt}}]{Jimenez:2022asc}%
  \BibitemOpen
  \bibfield  {author} {\bibinfo {author} {\bibfnamefont {R.}~\bibnamefont {Jimenez}}, \bibinfo {author} {\bibfnamefont {A.~R.}\ \bibnamefont {Khalifeh}}, \bibinfo {author} {\bibfnamefont {D.~F.}\ \bibnamefont {Litim}}, \bibinfo {author} {\bibfnamefont {S.}~\bibnamefont {Matarrese}}, \ and\ \bibinfo {author} {\bibfnamefont {B.~D.}\ \bibnamefont {Wandelt}},\ }\href {\doibase 10.1088/1475-7516/2023/09/007} {\bibfield  {journal} {\bibinfo  {journal} {JCAP}\ }\textbf {\bibinfo {volume} {09}},\ \bibinfo {pages} {007} (\bibinfo {year} {2023})},\ \Eprint {http://arxiv.org/abs/2210.10102} {arXiv:2210.10102 [astro-ph.CO]} \BibitemShut {NoStop}%
\bibitem [{\citenamefont {Sen}\ and\ \citenamefont {Sethi}(2002)}]{Sen_2002}%
  \BibitemOpen
  \bibfield  {author} {\bibinfo {author} {\bibfnamefont {A.~A.}\ \bibnamefont {Sen}}\ and\ \bibinfo {author} {\bibfnamefont {S.}~\bibnamefont {Sethi}},\ }\href {\doibase 10.1016/s0370-2693(02)01547-2} {\bibfield  {journal} {\bibinfo  {journal} {Physics Letters B}\ }\textbf {\bibinfo {volume} {532}},\ \bibinfo {pages} {159–165} (\bibinfo {year} {2002})}\BibitemShut {NoStop}%
\bibitem [{\citenamefont {Bento}\ \emph {et~al.}(2004)\citenamefont {Bento}, \citenamefont {Bertolami},\ and\ \citenamefont {Sen}}]{Bento_2004}%
  \BibitemOpen
  \bibfield  {author} {\bibinfo {author} {\bibfnamefont {M.~C.}\ \bibnamefont {Bento}}, \bibinfo {author} {\bibfnamefont {O.}~\bibnamefont {Bertolami}}, \ and\ \bibinfo {author} {\bibfnamefont {A.~A.}\ \bibnamefont {Sen}},\ }\href {\doibase 10.1103/physrevd.70.083519} {\bibfield  {journal} {\bibinfo  {journal} {Physical Review D}\ }\textbf {\bibinfo {volume} {70}} (\bibinfo {year} {2004}),\ 10.1103/physrevd.70.083519}\BibitemShut {NoStop}%
\bibitem [{\citenamefont {Adil}\ \emph {et~al.}(2024)\citenamefont {Adil}, \citenamefont {Akarsu}, \citenamefont {Di~Valentino}, \citenamefont {Nunes}, \citenamefont {\"Oz\"ulker}, \citenamefont {Sen},\ and\ \citenamefont {Specogna}}]{Adil_2024}%
  \BibitemOpen
  \bibfield  {author} {\bibinfo {author} {\bibfnamefont {S.~A.}\ \bibnamefont {Adil}}, \bibinfo {author} {\bibfnamefont {O.}~\bibnamefont {Akarsu}}, \bibinfo {author} {\bibfnamefont {E.}~\bibnamefont {Di~Valentino}}, \bibinfo {author} {\bibfnamefont {R.~C.}\ \bibnamefont {Nunes}}, \bibinfo {author} {\bibfnamefont {E.}~\bibnamefont {\"Oz\"ulker}}, \bibinfo {author} {\bibfnamefont {A.~A.}\ \bibnamefont {Sen}}, \ and\ \bibinfo {author} {\bibfnamefont {E.}~\bibnamefont {Specogna}},\ }\href {\doibase 10.1103/PhysRevD.109.023527} {\bibfield  {journal} {\bibinfo  {journal} {Phys. Rev. D}\ }\textbf {\bibinfo {volume} {109}},\ \bibinfo {pages} {023527} (\bibinfo {year} {2024})},\ \Eprint {http://arxiv.org/abs/2306.08046} {arXiv:2306.08046 [astro-ph.CO]} \BibitemShut {NoStop}%
\bibitem [{\citenamefont {Sahni}\ and\ \citenamefont {Sen}(2017)}]{Sahni:2015hbf}%
  \BibitemOpen
  \bibfield  {author} {\bibinfo {author} {\bibfnamefont {V.}~\bibnamefont {Sahni}}\ and\ \bibinfo {author} {\bibfnamefont {A.~A.}\ \bibnamefont {Sen}},\ }\href {\doibase 10.1140/epjc/s10052-017-4796-7} {\bibfield  {journal} {\bibinfo  {journal} {Eur. Phys. J. C}\ }\textbf {\bibinfo {volume} {77}},\ \bibinfo {pages} {225} (\bibinfo {year} {2017})},\ \Eprint {http://arxiv.org/abs/1510.09010} {arXiv:1510.09010 [astro-ph.CO]} \BibitemShut {NoStop}%
\bibitem [{\citenamefont {Alam}\ \emph {et~al.}(2021)\citenamefont {Alam} \emph {et~al.}}]{eBOSS:2020yzd}%
  \BibitemOpen
  \bibfield  {author} {\bibinfo {author} {\bibfnamefont {S.}~\bibnamefont {Alam}} \emph {et~al.} (\bibinfo {collaboration} {eBOSS}),\ }\href {\doibase 10.1103/PhysRevD.103.083533} {\bibfield  {journal} {\bibinfo  {journal} {Phys. Rev. D}\ }\textbf {\bibinfo {volume} {103}},\ \bibinfo {pages} {083533} (\bibinfo {year} {2021})},\ \Eprint {http://arxiv.org/abs/2007.08991} {arXiv:2007.08991 [astro-ph.CO]} \BibitemShut {NoStop}%
\bibitem [{\citenamefont {Nesseris}\ \emph {et~al.}(2017)\citenamefont {Nesseris}, \citenamefont {Pantazis},\ and\ \citenamefont {Perivolaropoulos}}]{Nesseris:2017vor}%
  \BibitemOpen
  \bibfield  {author} {\bibinfo {author} {\bibfnamefont {S.}~\bibnamefont {Nesseris}}, \bibinfo {author} {\bibfnamefont {G.}~\bibnamefont {Pantazis}}, \ and\ \bibinfo {author} {\bibfnamefont {L.}~\bibnamefont {Perivolaropoulos}},\ }\href {\doibase 10.1103/PhysRevD.96.023542} {\bibfield  {journal} {\bibinfo  {journal} {Phys. Rev. D}\ }\textbf {\bibinfo {volume} {96}},\ \bibinfo {pages} {023542} (\bibinfo {year} {2017})},\ \Eprint {http://arxiv.org/abs/1703.10538} {arXiv:1703.10538 [astro-ph.CO]} \BibitemShut {NoStop}%
\bibitem [{\citenamefont {Peirone}\ \emph {et~al.}(2017)\citenamefont {Peirone}, \citenamefont {Martinelli}, \citenamefont {Raveri},\ and\ \citenamefont {Silvestri}}]{Peirone:2017lgi}%
  \BibitemOpen
  \bibfield  {author} {\bibinfo {author} {\bibfnamefont {S.}~\bibnamefont {Peirone}}, \bibinfo {author} {\bibfnamefont {M.}~\bibnamefont {Martinelli}}, \bibinfo {author} {\bibfnamefont {M.}~\bibnamefont {Raveri}}, \ and\ \bibinfo {author} {\bibfnamefont {A.}~\bibnamefont {Silvestri}},\ }\href {\doibase 10.1103/PhysRevD.96.063524} {\bibfield  {journal} {\bibinfo  {journal} {Phys. Rev. D}\ }\textbf {\bibinfo {volume} {96}},\ \bibinfo {pages} {063524} (\bibinfo {year} {2017})},\ \Eprint {http://arxiv.org/abs/1702.06526} {arXiv:1702.06526 [astro-ph.CO]} \BibitemShut {NoStop}%
\bibitem [{\citenamefont {Raveri}\ \emph {et~al.}(2017)\citenamefont {Raveri}, \citenamefont {Bull}, \citenamefont {Silvestri},\ and\ \citenamefont {Pogosian}}]{Raveri:2017qvt}%
  \BibitemOpen
  \bibfield  {author} {\bibinfo {author} {\bibfnamefont {M.}~\bibnamefont {Raveri}}, \bibinfo {author} {\bibfnamefont {P.}~\bibnamefont {Bull}}, \bibinfo {author} {\bibfnamefont {A.}~\bibnamefont {Silvestri}}, \ and\ \bibinfo {author} {\bibfnamefont {L.}~\bibnamefont {Pogosian}},\ }\href {\doibase 10.1103/PhysRevD.96.083509} {\bibfield  {journal} {\bibinfo  {journal} {Phys. Rev. D}\ }\textbf {\bibinfo {volume} {96}},\ \bibinfo {pages} {083509} (\bibinfo {year} {2017})},\ \Eprint {http://arxiv.org/abs/1703.05297} {arXiv:1703.05297 [astro-ph.CO]} \BibitemShut {NoStop}%
\bibitem [{\citenamefont {Vagnozzi}\ \emph {et~al.}(2018)\citenamefont {Vagnozzi}, \citenamefont {Dhawan}, \citenamefont {Gerbino}, \citenamefont {Freese}, \citenamefont {Goobar},\ and\ \citenamefont {Mena}}]{Vagnozzi:2018jhn}%
  \BibitemOpen
  \bibfield  {author} {\bibinfo {author} {\bibfnamefont {S.}~\bibnamefont {Vagnozzi}}, \bibinfo {author} {\bibfnamefont {S.}~\bibnamefont {Dhawan}}, \bibinfo {author} {\bibfnamefont {M.}~\bibnamefont {Gerbino}}, \bibinfo {author} {\bibfnamefont {K.}~\bibnamefont {Freese}}, \bibinfo {author} {\bibfnamefont {A.}~\bibnamefont {Goobar}}, \ and\ \bibinfo {author} {\bibfnamefont {O.}~\bibnamefont {Mena}},\ }\href {\doibase 10.1103/PhysRevD.98.083501} {\bibfield  {journal} {\bibinfo  {journal} {Phys. Rev. D}\ }\textbf {\bibinfo {volume} {98}},\ \bibinfo {pages} {083501} (\bibinfo {year} {2018})},\ \Eprint {http://arxiv.org/abs/1801.08553} {arXiv:1801.08553 [astro-ph.CO]} \BibitemShut {NoStop}%
\bibitem [{\citenamefont {{Foreman-Mackey}}\ \emph {et~al.}(2013)\citenamefont {{Foreman-Mackey}}, \citenamefont {{Hogg}}, \citenamefont {{Lang}},\ and\ \citenamefont {{Goodman}}}]{Foreman-Mackey13}%
  \BibitemOpen
  \bibfield  {author} {\bibinfo {author} {\bibfnamefont {D.}~\bibnamefont {{Foreman-Mackey}}}, \bibinfo {author} {\bibfnamefont {D.~W.}\ \bibnamefont {{Hogg}}}, \bibinfo {author} {\bibfnamefont {D.}~\bibnamefont {{Lang}}}, \ and\ \bibinfo {author} {\bibfnamefont {J.}~\bibnamefont {{Goodman}}},\ }\href {\doibase 10.1086/670067} {\bibfield  {journal} {\bibinfo  {journal} {\pasp}\ }\textbf {\bibinfo {volume} {125}},\ \bibinfo {pages} {306} (\bibinfo {year} {2013})},\ \Eprint {http://arxiv.org/abs/1202.3665} {arXiv:1202.3665 [astro-ph.IM]} \BibitemShut {NoStop}%
\bibitem [{\citenamefont {{Lewis}}(2019)}]{Lewis:2019xzd}%
  \BibitemOpen
  \bibfield  {author} {\bibinfo {author} {\bibfnamefont {A.}~\bibnamefont {{Lewis}}},\ }\href@noop {} {\bibfield  {journal} {\bibinfo  {journal} {arXiv e-prints}\ ,\ \bibinfo {eid} {arXiv:1910.13970}} (\bibinfo {year} {2019})},\ \Eprint {http://arxiv.org/abs/1910.13970} {arXiv:1910.13970 [astro-ph.IM]} \BibitemShut {NoStop}%
\bibitem [{\citenamefont {Trotta}(2017)}]{Trotta:2017wnx}%
  \BibitemOpen
  \bibfield  {author} {\bibinfo {author} {\bibfnamefont {R.}~\bibnamefont {Trotta}}\ }(\bibinfo {year} {2017})\ \Eprint {http://arxiv.org/abs/1701.01467} {arXiv:1701.01467 [astro-ph.CO]} \BibitemShut {NoStop}%
\bibitem [{\citenamefont {Trotta}(2008)}]{Trotta:2008qt}%
  \BibitemOpen
  \bibfield  {author} {\bibinfo {author} {\bibfnamefont {R.}~\bibnamefont {Trotta}},\ }\href {\doibase 10.1080/00107510802066753} {\bibfield  {journal} {\bibinfo  {journal} {Contemp. Phys.}\ }\textbf {\bibinfo {volume} {49}},\ \bibinfo {pages} {71} (\bibinfo {year} {2008})},\ \Eprint {http://arxiv.org/abs/0803.4089} {arXiv:0803.4089 [astro-ph]} \BibitemShut {NoStop}%
\bibitem [{\citenamefont {Feroz}\ \emph {et~al.}(2009)\citenamefont {Feroz}, \citenamefont {Hobson},\ and\ \citenamefont {Bridges}}]{Feroz:2008xx}%
  \BibitemOpen
  \bibfield  {author} {\bibinfo {author} {\bibfnamefont {F.}~\bibnamefont {Feroz}}, \bibinfo {author} {\bibfnamefont {M.~P.}\ \bibnamefont {Hobson}}, \ and\ \bibinfo {author} {\bibfnamefont {M.}~\bibnamefont {Bridges}},\ }\href {\doibase 10.1111/j.1365-2966.2009.14548.x} {\bibfield  {journal} {\bibinfo  {journal} {Mon. Not. Roy. Astron. Soc.}\ }\textbf {\bibinfo {volume} {398}},\ \bibinfo {pages} {1601} (\bibinfo {year} {2009})},\ \Eprint {http://arxiv.org/abs/0809.3437} {arXiv:0809.3437 [astro-ph]} \BibitemShut {NoStop}%
\bibitem [{\citenamefont {Sahni}\ and\ \citenamefont {Shtanov}(2003)}]{Sahni:2002dx}%
  \BibitemOpen
  \bibfield  {author} {\bibinfo {author} {\bibfnamefont {V.}~\bibnamefont {Sahni}}\ and\ \bibinfo {author} {\bibfnamefont {Y.}~\bibnamefont {Shtanov}},\ }\href {\doibase 10.1088/1475-7516/2003/11/014} {\bibfield  {journal} {\bibinfo  {journal} {JCAP}\ }\textbf {\bibinfo {volume} {11}},\ \bibinfo {pages} {014} (\bibinfo {year} {2003})},\ \Eprint {http://arxiv.org/abs/astro-ph/0202346} {arXiv:astro-ph/0202346} \BibitemShut {NoStop}%
\bibitem [{\citenamefont {{Verde}}\ \emph {et~al.}(2023)\citenamefont {{Verde}}, \citenamefont {{Sch{\"o}neberg}},\ and\ \citenamefont {{Gil-Mar{\'\i}n}}}]{Verde:2023lmm}%
  \BibitemOpen
  \bibfield  {author} {\bibinfo {author} {\bibfnamefont {L.}~\bibnamefont {{Verde}}}, \bibinfo {author} {\bibfnamefont {N.}~\bibnamefont {{Sch{\"o}neberg}}}, \ and\ \bibinfo {author} {\bibfnamefont {H.}~\bibnamefont {{Gil-Mar{\'\i}n}}},\ }\href {\doibase 10.48550/arXiv.2311.13305} {\bibfield  {journal} {\bibinfo  {journal} {arXiv e-prints}\ ,\ \bibinfo {eid} {arXiv:2311.13305}} (\bibinfo {year} {2023})},\ \Eprint {http://arxiv.org/abs/2311.13305} {arXiv:2311.13305 [astro-ph.CO]} \BibitemShut {NoStop}%
\bibitem [{\citenamefont {Riess}\ \emph {et~al.}(2024)\citenamefont {Riess}, \citenamefont {Anand}, \citenamefont {Yuan}, \citenamefont {Casertano}, \citenamefont {Dolphin}, \citenamefont {Macri}, \citenamefont {Breuval}, \citenamefont {Scolnic}, \citenamefont {Perrin},\ and\ \citenamefont {Anderson}}]{Riess:2024ohe}%
  \BibitemOpen
  \bibfield  {author} {\bibinfo {author} {\bibfnamefont {A.~G.}\ \bibnamefont {Riess}}, \bibinfo {author} {\bibfnamefont {G.~S.}\ \bibnamefont {Anand}}, \bibinfo {author} {\bibfnamefont {W.}~\bibnamefont {Yuan}}, \bibinfo {author} {\bibfnamefont {S.}~\bibnamefont {Casertano}}, \bibinfo {author} {\bibfnamefont {A.}~\bibnamefont {Dolphin}}, \bibinfo {author} {\bibfnamefont {L.~M.}\ \bibnamefont {Macri}}, \bibinfo {author} {\bibfnamefont {L.}~\bibnamefont {Breuval}}, \bibinfo {author} {\bibfnamefont {D.}~\bibnamefont {Scolnic}}, \bibinfo {author} {\bibfnamefont {M.}~\bibnamefont {Perrin}}, \ and\ \bibinfo {author} {\bibfnamefont {I.~R.}\ \bibnamefont {Anderson}},\ }\href {\doibase 10.3847/2041-8213/ad1ddd} {\bibfield  {journal} {\bibinfo  {journal} {Astrophys. J. Lett.}\ }\textbf {\bibinfo {volume} {962}},\ \bibinfo {pages} {L17} (\bibinfo {year} {2024})},\ \Eprint {http://arxiv.org/abs/2401.04773} {arXiv:2401.04773 [astro-ph.CO]} \BibitemShut {NoStop}%
\bibitem [{\citenamefont {Di~Valentino}\ \emph {et~al.}(2020)\citenamefont {Di~Valentino} \emph {et~al.}}]{DiValentino:2020vvd}%
  \BibitemOpen
  \bibfield  {author} {\bibinfo {author} {\bibfnamefont {E.}~\bibnamefont {Di~Valentino}} \emph {et~al.},\ }\href@noop {} {\  (\bibinfo {year} {2020})},\ \Eprint {http://arxiv.org/abs/2008.11285} {arXiv:2008.11285 [astro-ph.CO]} \BibitemShut {NoStop}%
\bibitem [{\citenamefont {{Peirone}}\ \emph {et~al.}(2017)\citenamefont {{Peirone}}, \citenamefont {{Martinelli}}, \citenamefont {{Raveri}},\ and\ \citenamefont {{Silvestri}}}]{Peirone2017:Quint}%
  \BibitemOpen
  \bibfield  {author} {\bibinfo {author} {\bibfnamefont {S.}~\bibnamefont {{Peirone}}}, \bibinfo {author} {\bibfnamefont {M.}~\bibnamefont {{Martinelli}}}, \bibinfo {author} {\bibfnamefont {M.}~\bibnamefont {{Raveri}}}, \ and\ \bibinfo {author} {\bibfnamefont {A.}~\bibnamefont {{Silvestri}}},\ }\href {\doibase 10.1103/PhysRevD.96.063524} {\bibfield  {journal} {\bibinfo  {journal} {\prd}\ }\textbf {\bibinfo {volume} {96}},\ \bibinfo {eid} {063524} (\bibinfo {year} {2017})},\ \Eprint {http://arxiv.org/abs/1702.06526} {arXiv:1702.06526 [astro-ph.CO]} \BibitemShut {NoStop}%
\end{thebibliography}%

\appendix

\subsection{Figures and Table of constraints}\label{sec:figures}

\begin{figure*}
\begin{minipage}{0.59\linewidth}
\includegraphics[scale=0.3]{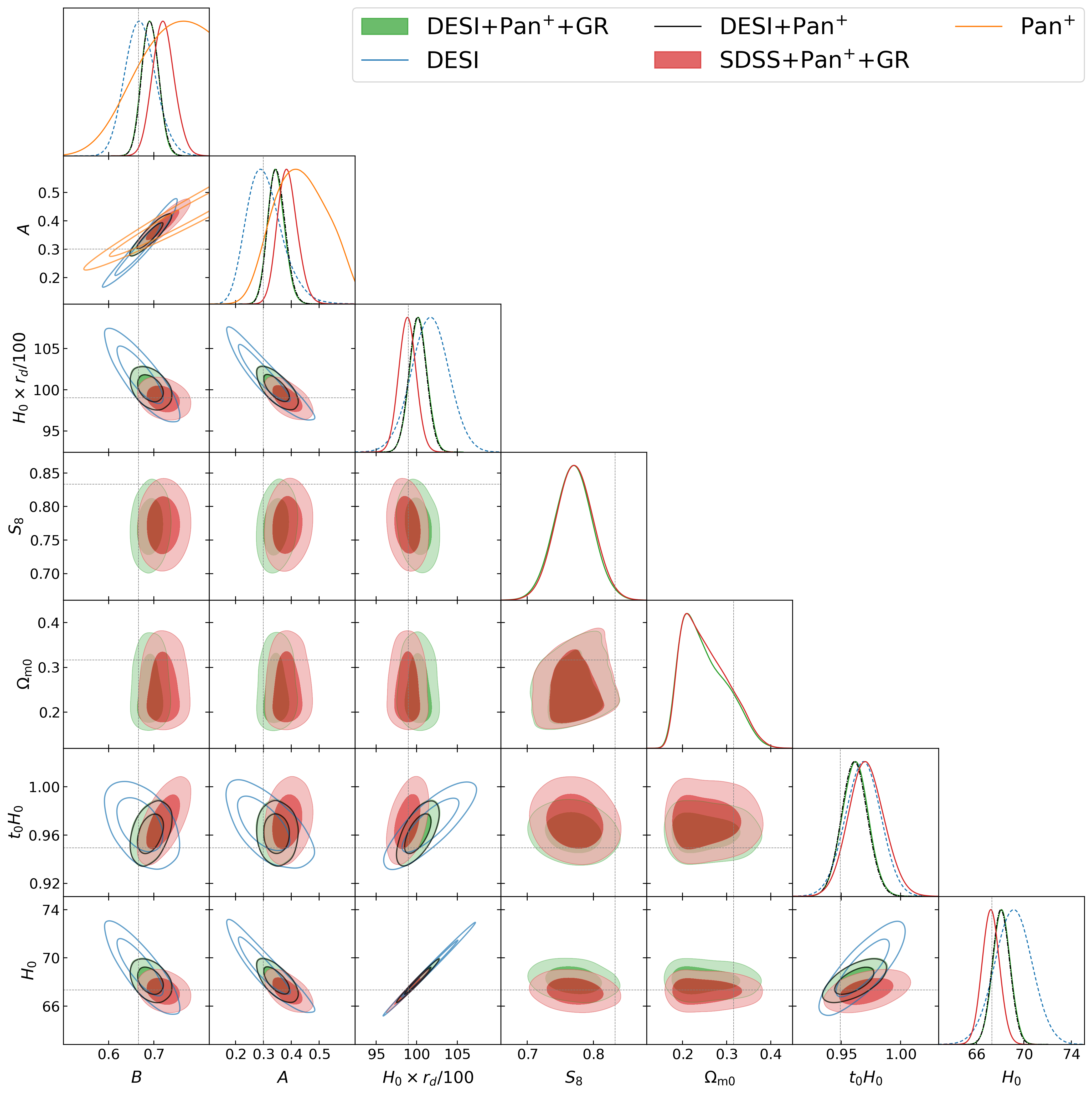}
\label{fig:prob1_6_2}
\end{minipage}
\begin{minipage}{0.39\textwidth}

\includegraphics[scale = 0.35]{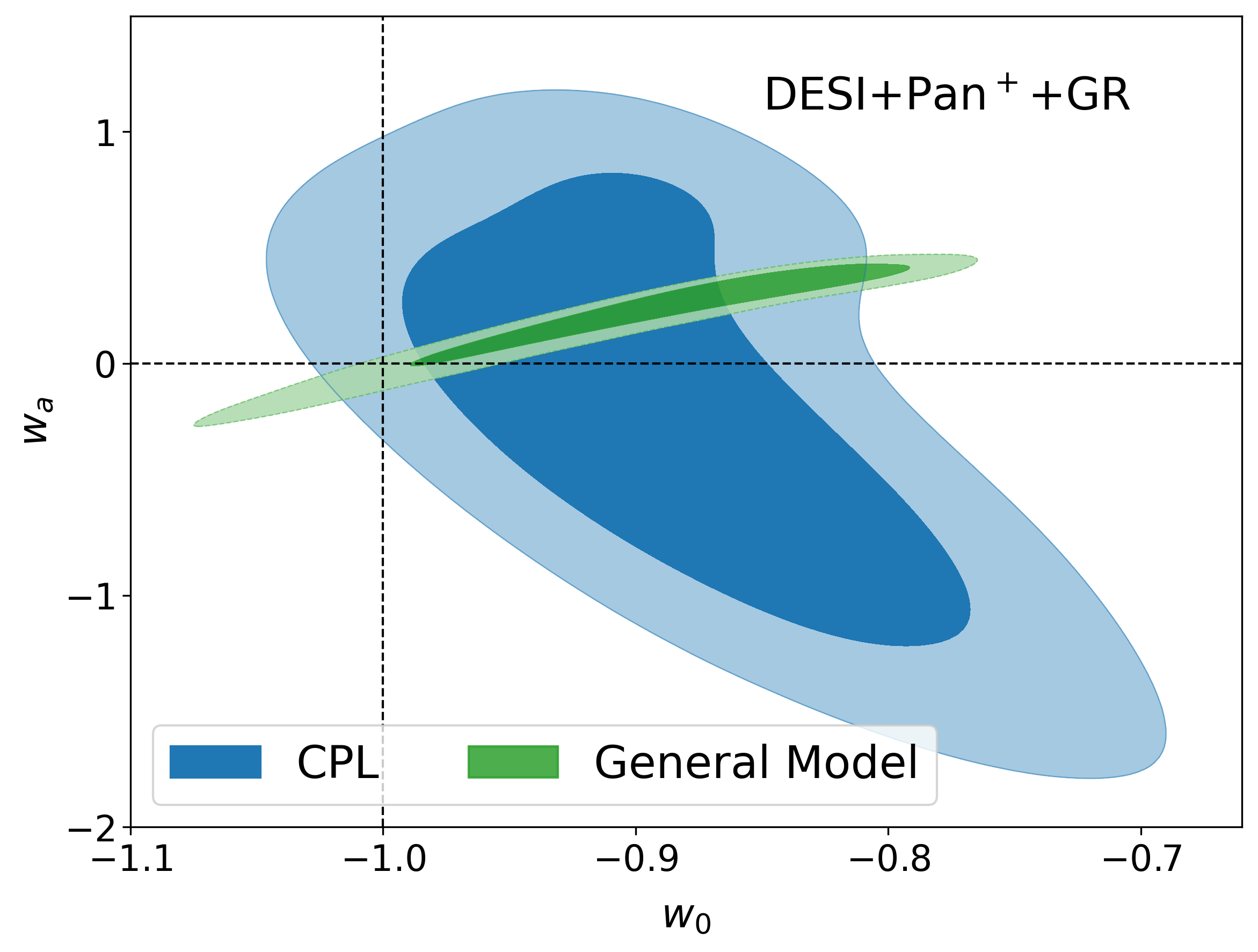}
\includegraphics[scale = 0.35]{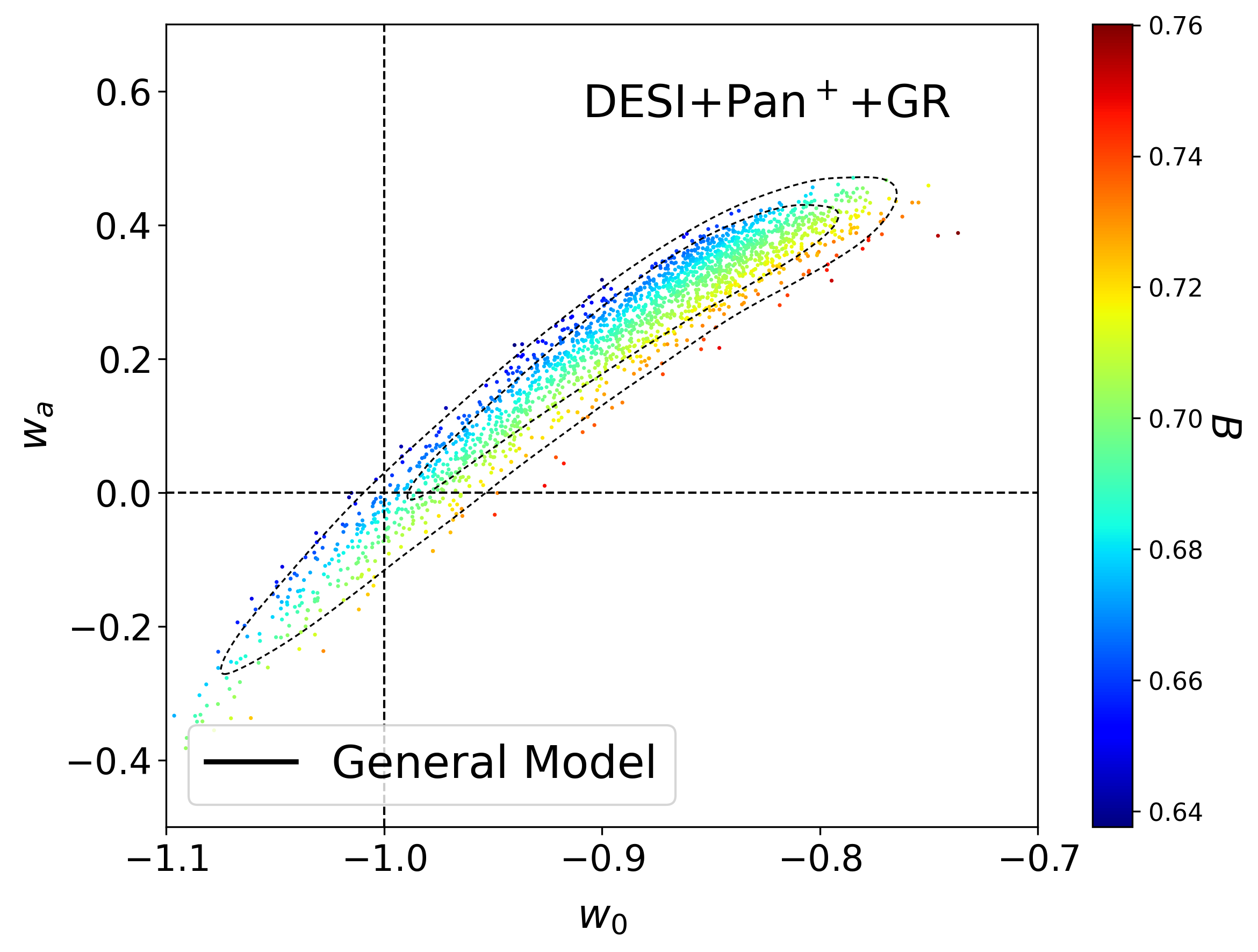}
\end{minipage}

\vspace{-0.1in}
\caption{\textit{Left}: Comparison of the constraints obtained for the general model using various datasets. We show the $68\%\, 95\%$ C.L. limits. We indicate the Planck-\LCDM model as markers in all the subplots. Here $\Hzero$ is derived in the inverse-distance-ladder approach utilising the prior $r_{\rm d} = 147.09 \pm 0.26 \, [\Mpc]$. \textit{Right-Top}: Comparison between CPL and the general model for the constrained parameters $w_0$, $w_a$ using DESI+Growth+\Panp data. \textit{Right-Bottom}: Same as the top panel but only for the general model, showing the distribution of the parameter $B$ in the  $w_0$, $w_a$ posteriors. } 
\label{triangle}\label{w0wa_BAOSNGRW}
\end{figure*}

For brevity in the main text, we show the contour plots and tables of constraints here. 

{\renewcommand{\arraystretch}{1.8}
\setlength{\tabcolsep}{6pt}

\begin{table*}[ht]

\begin{tabular}{ | m{1cm} || m{1.1cm}| m{1.3cm} | m{1.3cm} | m{1.3cm} | m{1.3cm} | m{1.5cm} | m{1.5cm} | m{1.5cm} | m{1.3cm} |}
 \hline
 \hline
 Model & Data & $A$ & $B$ & $\Omzero$ & $\sigma_8$ & $ h \times r_d$ & $w_0$ & $w_a$ & $\log(\mathcal Z)$\\ 
  \hline
  \hline
    & DESI & $0.31^{+0.05}_{-0.07}$ & $0.67^{+0.03}_{-0.04}$ & -- & -- & -- & -- & -- & --\\ 

    & \Panp & $0.44^{+0.09}_{-0.11}$ & $0.79^{+0.11}_{-0.12}$ & -- & -- & -- & -- & -- & --\\  

    General Model & DESI+ \Panp & $0.35^{+0.03}_{-0.03}$ & $0.69^{+0.02}_{-0.02}$ & -- & -- & $100.10^{+1.10}_{-1.10}$ & -- & -- & --\\

    & Joint-I & $0.35^{+0.03}_{-0.03}$ & $0.69^{+0.02}_{-0.02}$ & $0.25^{+0.04}_{-0.07}$ & $0.85^{+0.12}_{-0.08}$ & $100.20^{+1.10}_{-1.10}$ & $-0.90^{+0.09}_{-0.05}$ & $0.22^{+0.20}_{-0.07}$ & $-732.90$\\

    & Joint-II & $0.39^{+0.03}_{-0.04}$ & $0.72^{+0.02}_{-0.03}$ & $0.25^{+0.03}_{-0.07}$ & $0.85^{+0.12}_{-0.07}$ & $98.90^{+1.10}_{-1.10}$ & $-0.87^{+0.09}_{-0.05}$ & $0.23^{+0.19}_{-0.06}$ & $-729.50$\\  

   \hline
    CPL & Joint-I & -- & -- & $0.29^{+0.04}_{-0.01}$ & $0.79^{+0.02}_{-0.06}$ & $100.20^{+1.10}_{-1.10}$ & $-0.88^{+0.06}_{-0.08}$ & $-0.24^{+0.66}_{-0.66}$ & $-733.10$\\
    
    & Joint-II & -- & -- & $0.25^{+0.04}_{-0.04}$ & $0.86^{+0.07}_{-0.09}$ & $99.00^{+1.00}_{-1.00}$ & $-0.87^{+0.06}_{-0.05}$ & $0.37^{+0.45}_{-0.20}$ & $-730.40$\\
    \hline  
 
\end{tabular}
\caption{Marginalised $1\sigma$ C.L. limits of the parameters in each of the dataset combinations for the General Model. Also comparison between the General model and the CPL model for the constraints on the DE EoS parameters. Here datasets Joint-I and Joint-II represent the combinations DESI+Growth+\Panp and SDSS+Growth+\Panp, respectively. The last column shows the Bayesian evidence, where all the estimates have an uncertainty of $\sigma_{\log(\mathcal Z)} = 0.14$.}\label{Table_bestfit} 
\end{table*}
}

\end{document}